\documentclass[article,12pt,showkeys,showpacs] {revtex4}
\usepackage{amsfonts,graphicx, amsmath,longtable,float}
\begin{document}
\title{Spontaneous $\mathcal{PT}$ symmetry breaking in non-Hermitian coupled cavities array}
\author{Yan Xing}
\affiliation{Department of Physics, College of Science, Yanbian University, Yanji, Jilin 133002, People's Republic of China}
\author{Lu Qi}
\affiliation{Department of Physics, College of Science, Yanbian University, Yanji, Jilin 133002, People's Republic of China}
\author{Ji Cao}
\affiliation{Department of Physics, College of Science, Yanbian University, Yanji, Jilin 133002, People's Republic of China}
\author{Dong-Yang Wang}
\affiliation{Department of Physics, College of Science, Yanbian University, Yanji, Jilin 133002, People's Republic of China}
\author{Cheng-Hua Bai}
\affiliation{Department of Physics, College of Science, Yanbian University, Yanji, Jilin 133002, People's Republic of China}
\author{Hong-Fu Wang\footnote{E-mail: hfwang@ybu.edu.cn}}
\affiliation{Department of Physics, College of Science, Yanbian University, Yanji, Jilin 133002, People's Republic of China}
\author{Ai-Dong Zhu}
\affiliation{Department of Physics, College of Science, Yanbian University, Yanji, Jilin 133002, People's Republic of China}
\author{Shou Zhang}
\affiliation{Department of Physics, College of Science, Yanbian University, Yanji, Jilin 133002, People's Republic of China}

\begin{abstract}
We study the effects of the position of the passive and active cavities on the spontaneous parity-time ($\mathcal{PT}$) symmetry breaking behavior in non-Hermitian coupled cavities array model. We analyze and discuss the energy eigenvalue spectrums and $\mathcal{PT}$ symmetry in the topologically trivial and nontrivial regimes under three different cases in detail, i.e., the passive and active cavities are located at, respectively, the two end positions, the second and penultimate positions, and each position in coupled cavities array. The odevity of the number of cavities is further considered to check the effects of the non-Hermitian terms applied on the $\mathcal{PT}$ symmetric and asymmetric systems. We find that the position of the passive and active cavities has remarkable impacts on the spontaneous $\mathcal{PT}$ symmetry breaking behavior, and in each case the system exhibits distinguishable and novel spontaneous $\mathcal{PT}$ symmetry breaking characteristic, respectively. The effects of the non-Hermitian terms on the $\mathcal{PT}$ symmetric and asymmetric systems due to the odevity are comparatively different in the first case while qualitatively same in the second case.
\pacs{11.30.Er, 03.65.Vf, 42.50.Pq}
\keywords{$\mathcal{PT}$ symmetry, topology, coupled cavity}
\end{abstract}
\maketitle
\section{Introduction}
Since the discovery of topological insulators, the research for a topological state of the matters has attracted intense interest in past years in condensed-matter physics~\cite{001MC82304510,002XS83105711} and atomic, molecular, and optical physics~\cite{003JFGP83152311}. Topological insulators are characterized by a full insulating gap in the bulk and gapless edge or surface states which are protected by time-reversal symmetry. These modes have unique transport direction and are very robust to the disorder and perturbation. Owing to these novel properties, multifarious theoretical models and experimental schemes based on different physical systems, including cold atoms trapped in optical lattices~\cite{01YRKJI46262809,02MNEM303100513,03KLRMAI10822530312,04FJDXRSL9006363814,05AWYC1712500515} and open systems~\cite{06HMXX9205212215,07ZHWX9303212016,08HWX4645514}, have been proposed. As one of the simplest systems of one-dimensional (1D) topological insulators, the Su-Schrieffer-Heeger (SSH) model is a standard tight-binding model with spontaneous dimerization proposed by Su, Schrieffer, and Heeger to describe 1D polyacetylene~\cite{W42169879}. Despite its simplicity, it has attracted extensive studies in the past decades as it exhibits rich physical phenomena, such as topological soliton excitation, fractional charge, and nontrivial edge states~\cite{H21238880,R133398,A6078188,J8818040102}. Hence, it is interesting and worthwhile to simulate and map the SSH model based on different physical systems, such as graphene ribbons~\cite{09PDG8419545211}, \emph{p}-orbit optical ladder systems~\cite{10XEW4152313}, and off-diagonal bichromatic optical lattices~\cite{11SKS11018040313}. In addition, As one of the controllable and easily constructional quantum simulators, arrays of cavities feature the individual control and readout and the crucial advances in cavity quantum electrodynamics (QED) have turned this system into one of the leading platforms for the study of problems in condensed-matter physics~\cite{12I8615314,13M284906,14A285606,15F070200307,16D7603180507,17M252708}.

On the other hand, in traditional quantum mechanics, one of the fundamental axioms is that Hermitian operators certainly stand for physical observables in the Hilbert space so that real energy eigenvalues and the conservation of probability could be guaranteed~\cite{01R94}. However, Bender and Boettcher pointed out that a non-Hermitian Hamiltonian with $\mathcal{PT}$ symmetry can also possess a completely real energy eigenvalue spectrum in 1998~\cite{02CS80524398}. What's more interesting is that such a non-Hermitian Hamiltonian will undergo a spontaneous $\mathcal{PT}$ symmetry breaking transition. The system in the unbroken $\mathcal{PT}$ symmetry phase exhibits a completely real energy eigenvalue spectrum and all the eigenfunctions of the Hamiltonian are also the eigenfunctions of the $\mathcal{PT}$ operator, showing that all the eigenfunctions are $\mathcal{PT}$ symmetric. While in the spontaneous $\mathcal{PT}$ symmetry breaking phase, the energy eigenvalue spectrum becomes partially or completely complex and not all the eigenfunctions of the Hamiltonian have $\mathcal{PT}$ symmetry. Inspired by the extremely interesting property, many non-Hermitian $\mathcal{PT}$ symmetric Hamiltonians have drawn much attention in recent years and different kinds of $\mathcal{PT}$ symmetric systems have been investigated, including quantum field theories~\cite{03CDH7002500104}, open quantum systems~\cite{04I4215300109}, the Anderson models for disorder systems~\cite{05IB80289798,06J6316510801,07L4226520409}, the optical systems with complex refractive indices~\cite{08SUN10108040208,09AZY8204381810,10HDVIT10903390212,11S10312360109,12ZKRD10003040208,13XJHXQYC11024390213}, the Dirac Hamiltonians of topological insulators ~\cite{14YT8415310111}, topological systems~\cite{15BRS8906210214,16XTYP9201211615,17QBSLR9402211916}, tight-binding chain~\cite{18ORTB10303040209,19LZ8005210709,20YDMA8203010310,21LZ8103210910}, spin chain~\cite{22G8205240410,23CGXZ9001210314}, and so on. Furthermore, the rapid developments of photonic lattices and crystals have made it possible to experimentally realize non-Hermitian $\mathcal{PT}$ symmetric systems and have opened up avenues for experimental verification of these theorems~\cite{24RKDZ32263207,25KRDZ10010390408,26CKRDMD619210,27AGDRMVGD10309390209,28SBUHMAF10802410112,29ACMGDU48816712}.

In fact, any topological system will always interact with its nearby environment, which leads to the dissipative effects. A frequently-used and elegant way of describing interactions with environments on the stationary level is given by the application of non-Hermitian potentials~\cite{N2011}. On the other hand, it turned out that $\mathcal{PT}$ symmetry is a powerful concept to effectively describe the systems interacting with the environments in such a way that they experience balanced loss and gain. Thus, for a specific interacting process, viz., the topological system possesses $\mathcal{PT}$ symmetry, it is interesting and worthy to study and simulate how the topological property of system is affected by the presence of balanced non-Hermitian potentials and what the difference of the spontaneous $\mathcal{PT}$ symmetry breaking behavior between the topologically nontrivial and trivial regimes. Very recently, there has been growing interest in $\mathcal{PT}$ symmetric or non-Hermitian 1D topological models. Zhu {\it et al}. have studied the $\mathcal{PT}$ symmetry in the non-Hermitian SSH model with two conjugated purely imaginary potentials at the two end sites~\cite{15BRS8906210214}. The spontaneous $\mathcal{PT}$ symmetry breaking in the non-Hermitian Kitaev and extended Kitaev models with two conjugated purely imaginary potentials at the two end sites also has been mentioned in Ref.~\cite{16XTYP9201211615}. Moreover, Zeng {\it et al}. have extensively discussed the effects of non-Hermitian terms on the nontrivial phase and the robustness of Majorana bound states in four kinds of generalized non-Hermitian Kitaev chain with imaginary potentials added to some or all the lattice sites~\cite{17QBSLR9402211916}. Although the topological insulators, cavity QED, and $\mathcal{PT}$ symmetry have been rapidly developed and extensively investigated, respectively, the connection among them has been less explored yet. It is the purpose of this work to provide an effective approach to study and simulate the SSH model and interacting process between the SSH model and its nearby environments based on cavity QED by utilizing the passive and active cavities~\cite{30JRY9205383715}, which is more realistic. Additionally, $\mathcal{PT}$ symmetric SSH model and spontaneous $\mathcal{PT}$ symmetry breaking of the $\mathcal{PT}$ symmetric SSH model can be further realized via adjusting balanced loss and gain.

To this end, we propose a scheme to investigate the spontaneous $\mathcal{PT}$ symmetry breaking behavior in non-Hermitian coupled cavities array model which is constructed by introducing additional passive and active cavities. The odevity of the number of cavities is also considered to check the effects of the non-Hermitian terms applied on the $\mathcal{PT}$ symmetric and asymmetric systems. By assigning alternatingly modulative coupling strength in non-Hermitian coupled cavities array, the model can be accurately mapped to a non-Hermitian SSH model with complex on-site potentials. The Hamiltonian of the system in the situation of even number of cavities satisfies the $\mathcal{PT}$ symmetry albeit do not obey $\mathcal{P}$ and $\mathcal{T}$ symmetries separately, while in the situation of odd number of cavities, it is not $\mathcal{PT}$ symmetric. We mainly discuss three different cases and find that if the passive and active cavities are located at the two end positions, for even number of cavities, the $\mathcal{PT}$ symmetry in the topologically nontrivial regime is spontaneously broken for an arbitrary nonzero effective loss rate $\kappa$. While in the topologically trivial regime, the system will undergo an abrupt transition from the unbroken $\mathcal{PT}$ symmetry phase to the spontaneous $\mathcal{PT}$ symmetry breaking phase at a critical value $\kappa_{c}$ and a second transition at another critical value $\kappa_{c^{'}}$. The total system exhibits complex energy eigenvalues once the effective loss rate $\kappa$ is nonzero except the two ``Dirac points" which exist entirely real energy eigenvalues and undergoes another transition when the number of cavities is odd. However, if the passive and active cavities are located at the second and penultimate positions, for even number of cavities, the system in the topologically nontrivial regime can exhibit an unbroken $\mathcal{PT}$ symmetry phase when $\kappa\leq\kappa_{c}$. Furthermore, all of the phase regimes will undergo the spontaneous $\mathcal{PT}$ symmetry breaking transition and a second transition at, respectively, critical values $\kappa_{c}$ and $\kappa_{c^{'}}$. For odd number of cavities, the total system exhibits similar behaviors compared with the case of even number of cavities. In a more general case of even number sequence of the passive and active cavities, as $\kappa$ ceaselessly increasing, the $\mathcal{PT}$ symmetry of the total system is spontaneously broken accompanying with the occurrence of large-scale purely imaginary energy eigenvalues. In the end, the whole energy eigenvalue becomes purely imaginary and the system exhibits a purely imaginary energy eigenvalue spectrum.

The organization of the paper is as follows. In Section \uppercase\expandafter{\romannumeral2}, the Hamiltonian of non-Hermitian coupled cavity arrays model is presented. In Section \uppercase\expandafter{\romannumeral3}, the energy eigenvalue spectrum, the spontaneous $\mathcal{PT}$ symmetry breaking behavior in different even cavity situations, the effects of the non-Hermitian terms on the $\mathcal{PT}$ symmetric and asymmetric systems due to the odevity, and the experimental feasibility are given and discussed. Finally, we summarize our results in Section \uppercase\expandafter{\romannumeral4}.

\section{Model Hamiltonian}
\begin{figure}
\includegraphics[scale=0.72]{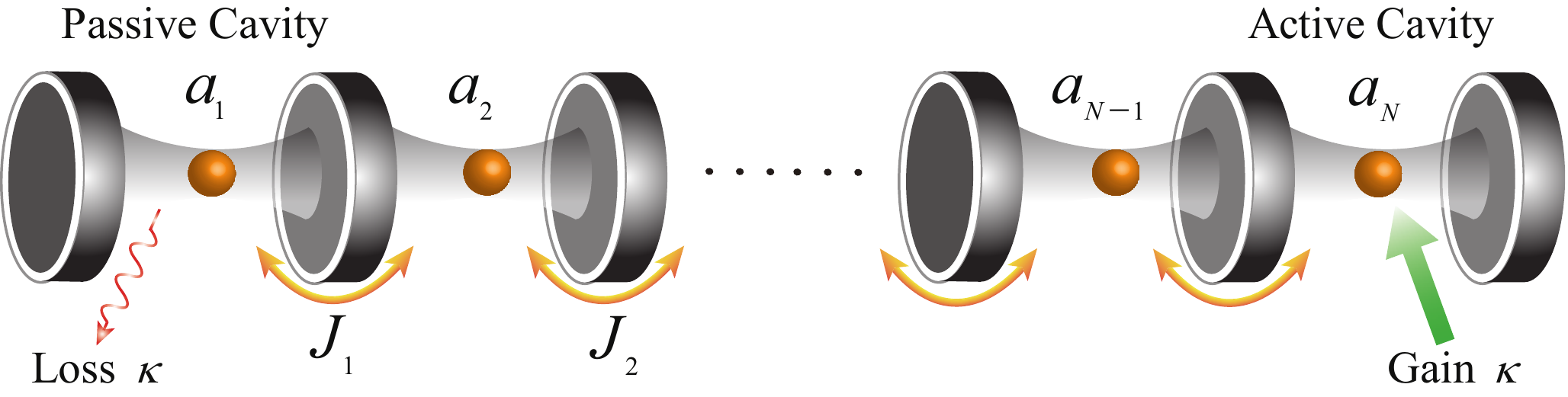}\label{fig1}
\centering
\caption{(Color online) Schematic illustration of non-Hermitian coupled cavities array model with alternatingly modulative coupling strength including an additional passive cavity at the first position and an additional active cavity at the $\emph{N}$th position, where a two-level atom is trapped in each cavity.}
\end{figure}

We consider a model composed of an array of cavities that are coupled via exchange of photons with one two-level atom in each cavity, as shown in Fig.~1. The Hamiltonian of the system is written as (setting $\hbar = 1$)
\begin{eqnarray}\label{e01}
H=H_{ac}+H_{cc},
\end{eqnarray}
with
\begin{eqnarray}\label{e02}
H_{ac}&=&\sum_{i=1}^{\emph{N}}\left(\frac{\omega_{a}}{2}\sigma_{i}^{z}+\omega_{c}a_{i}^{\dag}a_{i}+
ga_{i}\sigma_{i}^{+}+ga_{i}^{\dag}\sigma_{i}^{-}\right),\cr\cr
H_{cc}&=&\sum_{j=2n-1}^{N-1}J_{1}a_{j}^{\dag}a_{j+1}+\sum_{j=2n}^{N-1}J_{2}a_{j}^{\dag}a_{j+1}+\mathrm{H.c.},
\end{eqnarray}
where \emph{N} is the total number of cavities, $a_{i}$ ($a_{i}^{\dag}$) is the annihilation (create) operator of the $i$th cavity mode, $\omega_{a}$ and $\omega_{c}$ are the frequencies of atom and cavity mode, $\sigma_{i}^{z}$, $\sigma_{i}^{+}$, and $\sigma_{i}^{-}$ are atomic operators, $g$ is the coupling constant between atom and cavity mode, respectively. The cavity-cavity hopping strengths are alternately modulative. Specifically, the coupling strength between cavities $j$ and $j+1$ is $J_{1}$ when $j$ is odd, whereas when $j$ is even, the coupling strength is $J_{2}$. In the rotating frame with respect to the external driving frequency $\omega_{d}$ and in the interaction picture with respect to the atomic frequency $\omega_{a}$, if all the atoms are prepared in the ground states, we can obtain the effective Hamiltonian as below
\begin{eqnarray}\label{e03}
H_{\mathrm{eff}}=\sum_{i=1}^{\emph{N}}\left(\Delta_{c}-\frac{g^{2}}{\Delta_{a}}\right)a_{i}^{\dag}a_{i}+\left(\sum_{j=2n-1}^{N-1}J_{1}a_{j}^{\dag}a_{j+1}+\sum_{j=2n}^{N-1}J_{2}a_{j}^{\dag}a_{j+1}+\mathrm{H.c.}\right),
\end{eqnarray}
where $\Delta_{c}=\omega_{c}-\omega_{d}$ ($\Delta_{a}=\omega_{a}-\omega_{d}$) is the detuning of cavity mode (atom) frequency from the driving field. Taking the spontaneous energy loss at the first cavity and energy gain at the last cavity into account, which means that a passive cavity and an active cavity are located at the two end positions of the coupled cavity arrays, respectively. The total effective non-Hermitian Hamiltonian takes the form
\begin{eqnarray}\label{e04}
H_{\mathrm{total}}&=&\sum_{i=1}^{\emph{N}}\left(\Delta_{c}-\frac{g^{2}}{\Delta_{a}}\right)a_{i}^{\dag}a_{i}-i\kappa_{1} a_{1}^{\dag}a_{1}-i\kappa_{\emph{N}}a_{\emph{N}}^{\dag}a_{\emph{N}}
\cr\cr&&+\left(\sum_{j=2n-1}^{N-1}J_{1}a_{j}^{\dag}a_{j+1}+\sum_{j=2n}^{N-1}J_{2}a_{j}^{\dag}a_{j+1}+\mathrm{H.c.}\right),
\end{eqnarray}
where $\kappa_{1}=\kappa_1^i+\kappa_1^e$ is the total loss rate of the passive cavity 1, with $\kappa_1^i$ being the intrinsic loss rate and $\kappa_1^e$ being the external coupling loss rate. In the active cavity $N$, on the other hand, the effective loss rate $\kappa_{N}=\kappa_N^i-\xi$ is reduced by the gain $\xi$ (round-trip energy gain). Here $\kappa_{N}>0$ (loss) corresponding to a passive cavity or $\kappa_{N}<0$ (gain) corresponding to an active cavity depends on $\xi$, which has been realized in recent experiments fortunately~\cite{31LXSCJLGGM852414,32BSFFMGSFCL1039414}.

Setting $\Delta_{c}-\frac{g^{2}}{\Delta_{a}}=\varepsilon$, $\kappa_{1}=-\kappa_{\emph{N}}=\kappa$ ($\kappa>0$), and the following parameter conditions:
\begin{eqnarray}\label{e05}
J_{1}&=&J\left(1-\delta\cos\Phi\right),\cr\cr
J_{2}&=&J\left(1+\delta\cos\Phi\right),
\end{eqnarray}
where the parameter $\Phi$ is a cyclical parameter which can vary from 0 to 2$\pi$ continuously and $\delta$ is the strength of cycle modulation. For convenience, $J = 1$ is set as the unit of energy, the Hamiltonian of the system thus can be rewritten as
\begin{eqnarray}\label{e06}
H_{\mathrm{s}}&=&\left[\sum_{j=2n-1}^{N-1}\left(1-\delta\cos\Phi\right)a_{j}^{\dag}a_{j+1}+\sum_{j=2n}^{N-1}\left(1+\delta\cos\Phi\right)a_{j}^{\dag}a_{j+1}
+\mathrm{H.c.}\right]\cr\cr
&&+\sum_{i=1}^{\emph{N}}\varepsilon a_{i}^{\dag}a_{i}-i\kappa a_{1}^{\dag}a_{1}+i\kappa a_{\emph{N}}^{\dag}a_{\emph{N}}.
\end{eqnarray}
The above Hamiltonian can be proved to be $\mathcal{PT}$ symmetric when the number of cavities is even, while it is not $\mathcal{PT}$ symmetric in the situation of odd number of cavities.

\section{RESULTS AND DISCUSSIONS}
\begin{figure}
\includegraphics[scale=0.8]{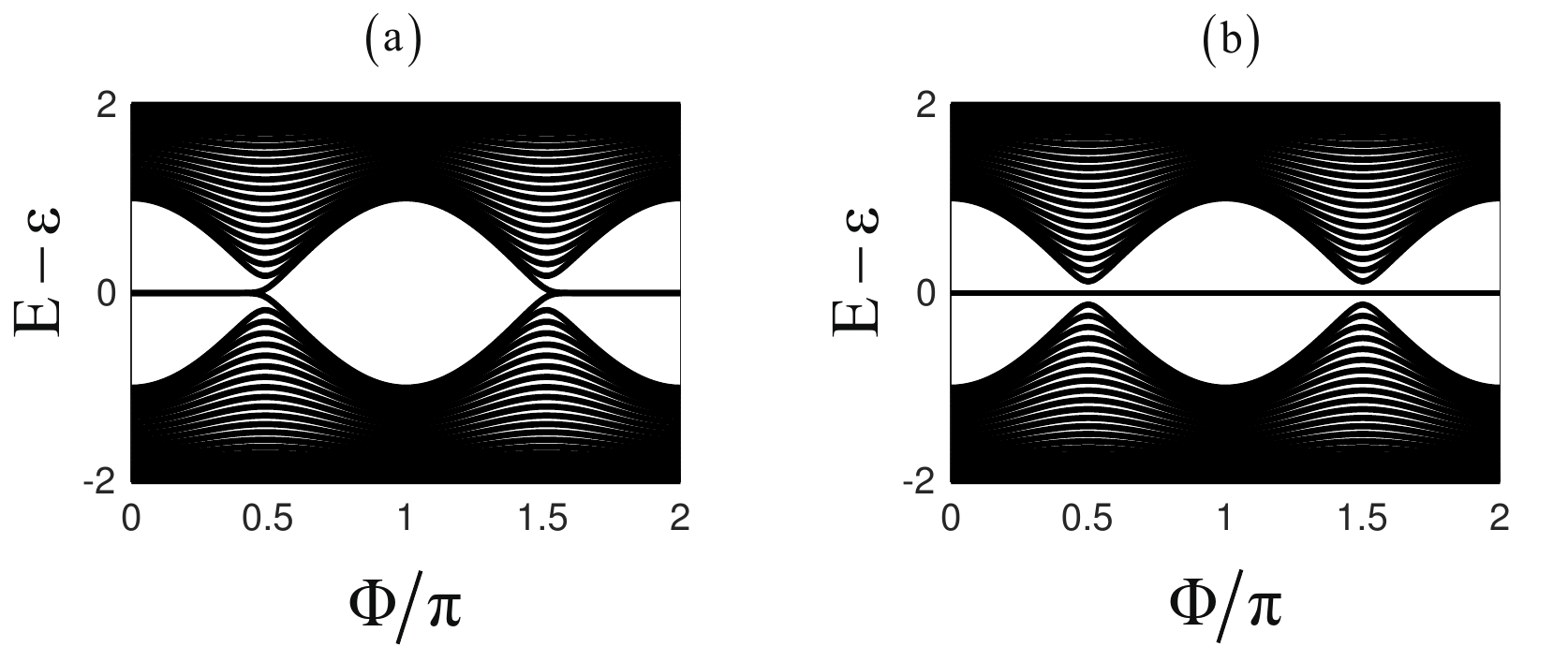}\label{fig2}
\centering
\caption{Energy eigenvalue spectrum for the Hermitian coupled cavities array model with parameters $\delta=0.5$ when (a) $\emph{N}=50$ and (b) $\emph{N}=51$.}
\end{figure}

In this section, we present the energy eigenvalue spectrum of the non-Hermitian coupled cavities array model and discuss the effects of the position of the passive and active cavities on the spontaneous $\mathcal{PT}$ symmetry breaking behavior of the system. Further we consider the odevity of the number of cavities to check the effects of the non-Hermitian terms applied on the $\mathcal{PT}$ symmetric and asymmetric systems. In the meanwhile, a more general $\mathcal{PT}$ symmetric case is presented. Without loss of generality, before proceeding, in the absence of the passive and active cavities, we plot the energy eigenvalue spectrum of the Hermitian system as functions of $\Phi$, which exhibits that the energy eigenvalue spectrum is analogous to the conventional SSH model essentially, as shown in Fig.~2. For even number cavities, in the regimes of $0<\Phi<\frac{\pi}{2}$ and $\frac{3\pi}{2}<\Phi<2\pi$, it is featured by the presence of twofold-degenerate zero-energy edge modes in the topologically nontrivial regime. Conversely, corresponding to the topologically trivial regime $\frac{\pi}{2}<\Phi<\frac{3\pi}{2}$, the system does not support the topologically nontrivial zero-energy edge modes. Remarkably, the bulk gap closes and reopens at the phase boundary points $\Phi=\frac{\pi}{2},\frac{3\pi}{2}$, as shown in Fig.~2(a). On the other hand, a single zero-energy mode will always emerge for all $\Phi$ when the number of cavities is odd, as shown in Fig.~2(b). This is the typically even-odd effect of the SSH model owing to the chiral symmetry.

\subsection{The passive and active cavities at the two end positions}
We first consider the situation that the number of cavities is even. The system is $\mathcal{PT}$ symmetric under this circumstance and numerical results of the energy eigenvalue spectrum for the $\mathcal{PT}$ symmetric coupled cavities array model governed by Hamiltonian~({\ref{e06}}) are shown in Fig.~3 and Fig.~4. Figure~3 shows the real and imaginary parts of the energy eigenvalue spectrum as functions of $\Phi$ for different $\kappa$. To begin with, we consider the topologically nontrivial regimes $0<\Phi<\frac{\pi}{2}$ and $\frac{3\pi}{2}<\Phi<2\pi$. When the effective loss rate $\kappa$ is weak, for example, $\kappa=0.1$, complex energy eigenvalues emerge in this regime, implying that the $\mathcal{PT}$ symmetry is spontaneously broken. As a matter of fact, we find that the system exhibits a pair of conjugated purely imaginary energy eigenvalues with the form of $\pm ib$ ($b$ is a function of $\Phi$ and $\kappa$) and $N$-$2$ real energy eigenvalues as long as $\kappa$ is nonzero, as shown in Fig.~3(a). With $\kappa$ continuously increasing, one can observe that the energy eigenvalue spectrum in the topologically nontrivial regime can also exist only a pair of conjugated purely imaginary energy eigenvalues, as shown in Figs.~3(b)$-$(e).

\begin{figure}
\includegraphics[scale=0.7]{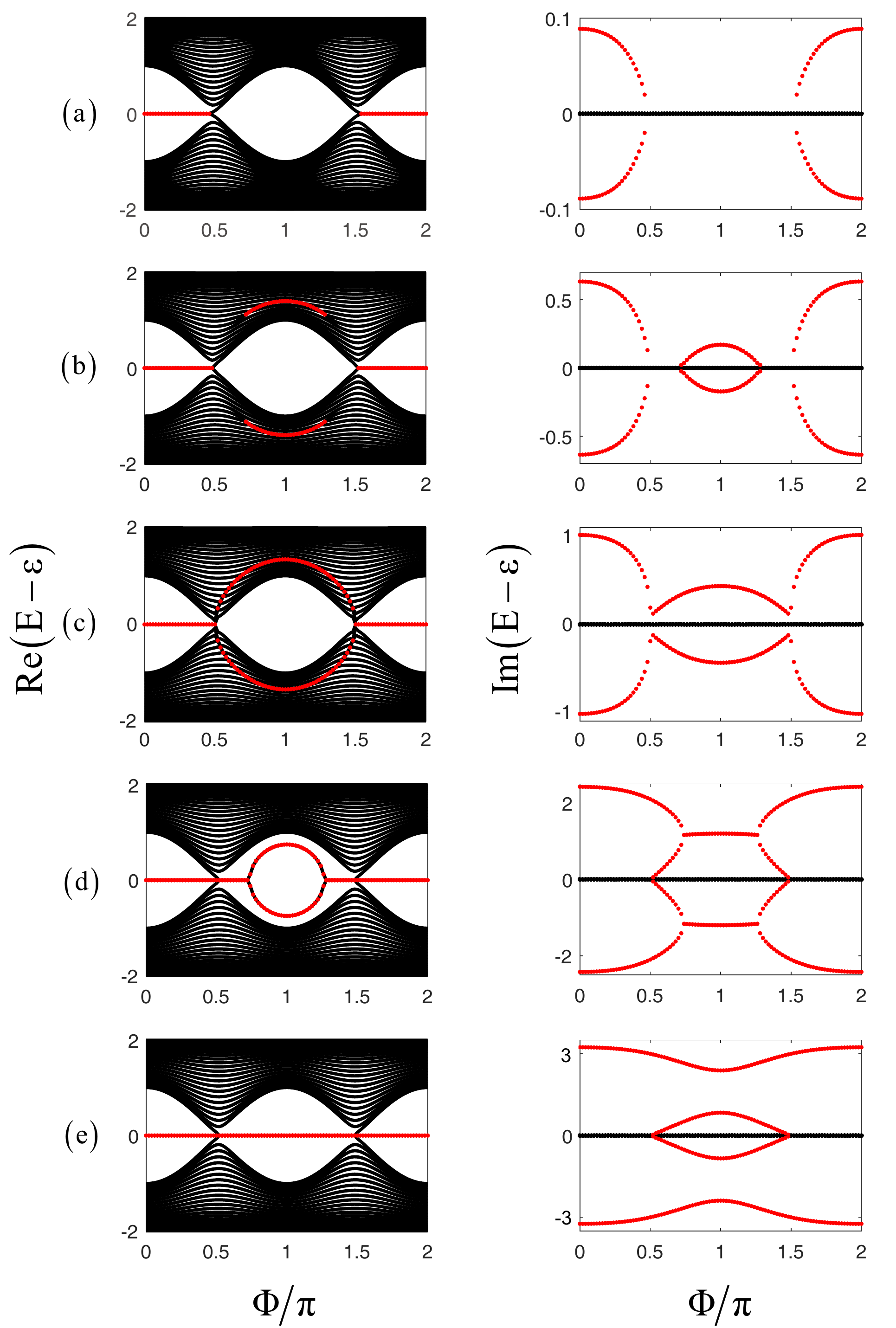}\label{fig3}
\centering
\caption{(Color online) The real and the imaginary parts of the energy eigenvalue spectrum of the $\mathcal{PT}$ symmetric coupled cavities array model as functions of $\Phi$ with parameters $\delta=0.5$ and $\emph{N}=50$ for different $\kappa$. (a) $\kappa=0.1$, (b) $\kappa=0.7$, (c) $\kappa=1.1$, (d) $\kappa=2.5$, and (e) $\kappa=3.3$. Left and right figures represent the real and imaginary parts of the energy eigenvalue spectrum, respectively. The red points represent the real and imaginary parts of the complex energy eigenvalues.}
\end{figure}

\begin{figure}
\includegraphics[scale=0.6]{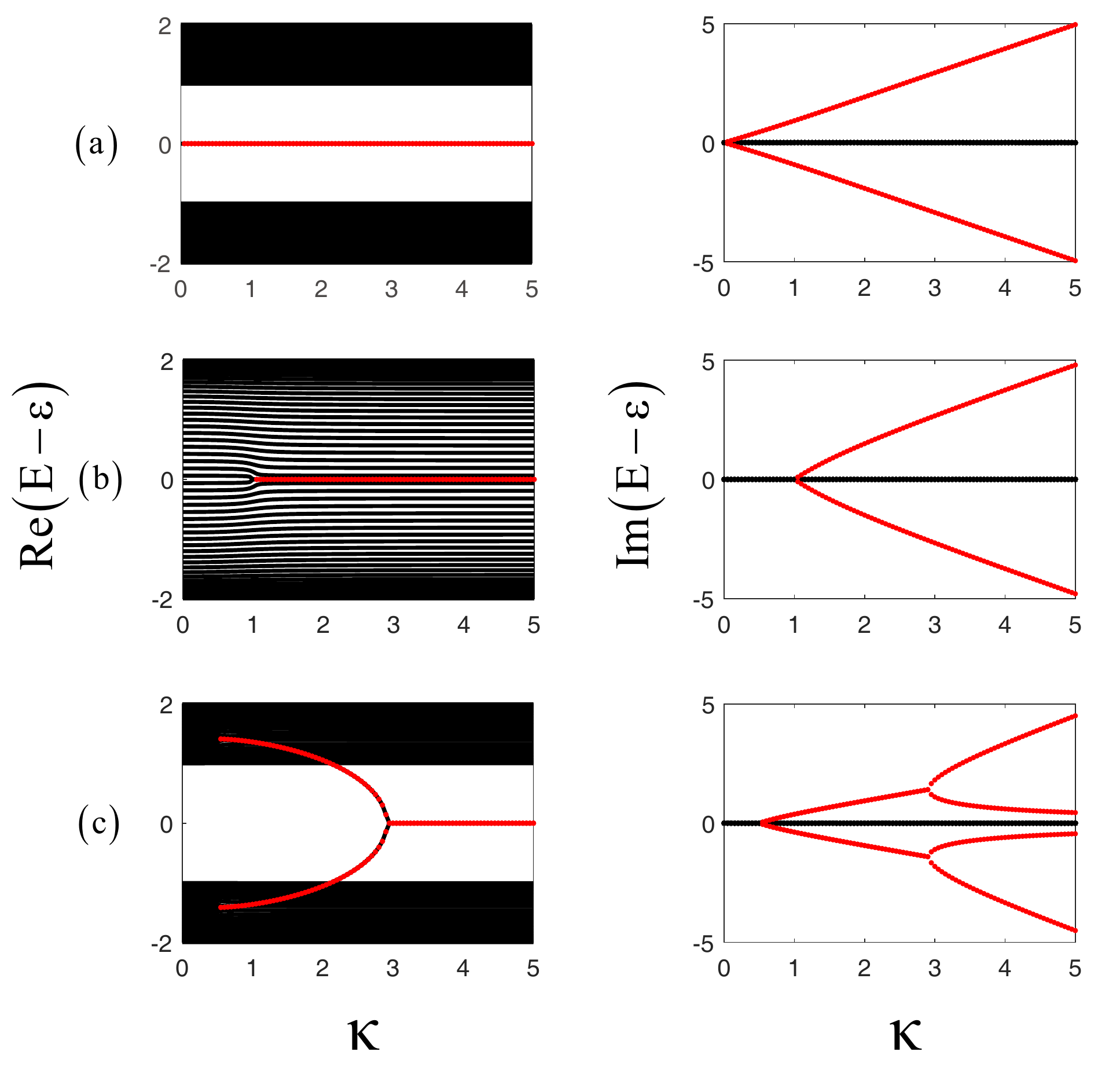}\label{fig4}
\centering
\caption{(Color online) The real and imaginary parts of the energy eigenvalue spectrum versus $\kappa$ for the system in different topological regimes. (a) System in the topologically nontrivial regime with $\Phi=0$, (b) system on the phase boundary points with $\Phi=\frac{\pi}{2}$, and (c) system in the topologically trivial regime with $\Phi=\pi$. Other
parameters are the same as Fig.~3. The red points represent the real and imaginary parts of the complex energy eigenvalues.}
\end{figure}

\begin{figure}
\centering
\includegraphics[scale=0.8]{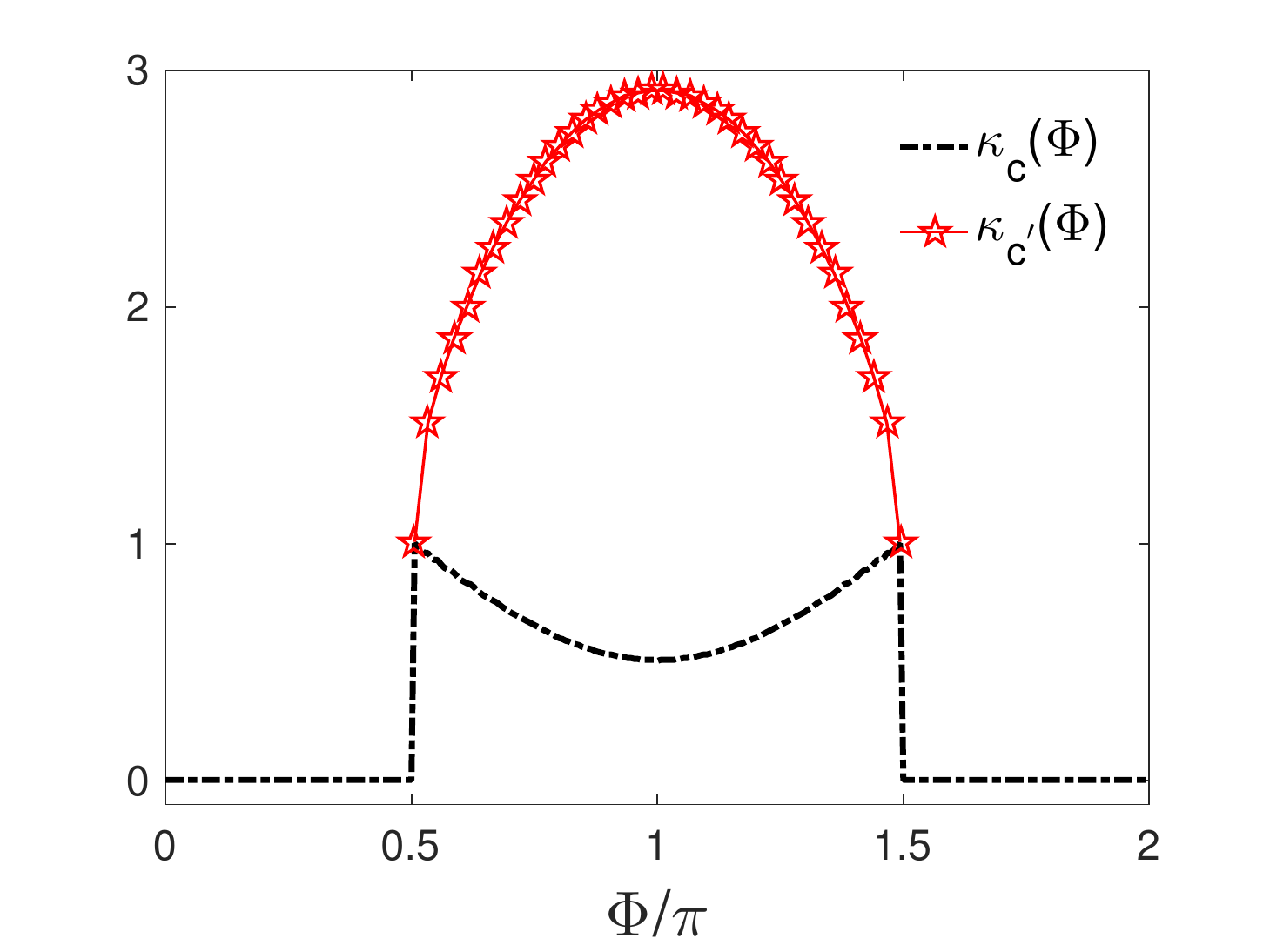}\label{fig5}
\caption{(Color online) The sketches of $\kappa_{c}(\Phi)$ and $\kappa_{c^{'}}(\Phi)$ for the system as function of $\Phi$.}
\end{figure}

In the topologically trivial regime $\frac{\pi}{2}<\Phi<\frac{3\pi}{2}$, it turns out that the system exhibits much richer characteristics, which have significant differences from the topologically nontrivial regime. In the case of weak effective loss rate $\kappa=0.1$, as shown in Fig.~3(a), the system has an entirely real energy eigenvalue spectrum, which indicates that the $\mathcal{PT}$ symmetry of the regime is unbroken. As $\kappa$ continues to increase, we find that the four complex energy eigenvalues with the form of $\pm a\pm ib$ begin to arise if $\kappa$ is larger than a critical value $\kappa_{c}$ (here $\kappa_{c}$ is a function of $\Phi$), which indicates that the system in this regime undergoes a spontaneous $\mathcal{PT}$ symmetry breaking transition at the critical value $\kappa_{c}$. The spontaneous $\mathcal{PT}$ symmetry breaking transition initially occurs at $\Phi=\pi$ with $\kappa_{c}(\pi)=0.502$ and the complex energy eigenvalues will extend from $\Phi=\pi$ to the phase boundary points with the increase of $\kappa$. It is worth mentioning that the system in this regime still has unbroken $\mathcal{PT}$ symmetry for a suitable $\kappa$, as shown in Fig.~3(b).

Figure~3(c) shows the $\mathcal{PT}$ symmetry breaking at the phase boundary points ($\pi/2$ and $3\pi/2$) with $\kappa_{c}(\frac{\pi}{2})=\kappa_{c}(\frac{3\pi}{2})=1$, in this case the $\mathcal{PT}$ symmetry of the total system is spontaneously broken. What's more interesting is that when $\kappa>1$, a novel behavior appears in the topologically trivial regime near the phase boundary points, which can be characterized by the split of the imaginary parts of the energy eigenvalues and the behavior spreads from the phase boundary points to $\Phi=\pi$ with a second critical value $\kappa_{c^{'}}(\Phi)$, corresponding to that four complex energy eigenvalues turn into two pairs of conjugated purely imaginary energy eigenvalues, as shown in Fig.~3(d). Further increasing $\kappa$, the whole topologically trivial regime exhibits two pairs of conjugated purely imaginary energy eigenvalues when $\kappa>\kappa_{c^{'}}(\pi)=2.91$, as shown in Fig.~3(e).

To illustrate these phenomena mentioned above more clearly, we plot the real and imaginary parts of the energy eigenvalue spectrum as functions of $\kappa$ for the system in different phase regimes in Fig.~4. As an example, the energy eigenvalue spectrum versus $\kappa$ for the system with $\Phi=0$ is given in Fig.~4(a) and it is obvious that the complex energy eigenvalues turn up in the topologically nontrivial regime once $\kappa\neq0$. Specifically, for the system with $\Phi=\frac{\pi}{2},\pi$, the energy eigenvalue spectrums as functions of $\kappa$ are also shown in Figs.~4(b)$-$(c). It is clear that the spontaneous $\mathcal{PT}$ symmetry breaking transition lastly takes place at the phase boundary points. Moreover, there exists unbroken $\mathcal{PT}$ symmetry for $\kappa\leq\kappa_{c}(\Phi)$ and the system reveals a first $\mathcal{PT}$ symmetry breaking transition and a second transition at a certain $\kappa_{c}(\Phi)$ and $\kappa_{c^{'}}(\Phi)$ in the topologically trivial regime, respectively. A pair of conjugated purely imaginary energy eigenvalues tend to zero corresponding to the two central red lines in Fig.~3(e) in the limit of $\kappa\rightarrow\infty$, as shown in Fig.~4(c). On the other hand, the approximate relationships between $\kappa_{c}(\Phi)$, $\kappa_{c^{'}}(\Phi)$ and $\Phi$, respectively, are also plotted in Fig.~5. From the numerical results we can draw the conclusion that the $\mathcal{PT}$ symmetry in the topologically nontrivial regime is spontaneously broken once $\kappa\neq0$. While in the topologically trivial regime the system exhibits an entirely real energy eigenvalue spectrum when $\kappa\leq\kappa_{c}(\Phi)$ and will undergo a spontaneous $\mathcal{PT}$ symmetry breaking transition and a second transition at a certain $\kappa_{c}(\Phi)$ and $\kappa_{c^{'}}(\Phi)$, respectively, which have an opposite transition direction. Furthermore, the $\mathcal{PT}$ symmetry at the phase boundary points is the most stable.

\begin{figure}
\includegraphics[scale=0.7]{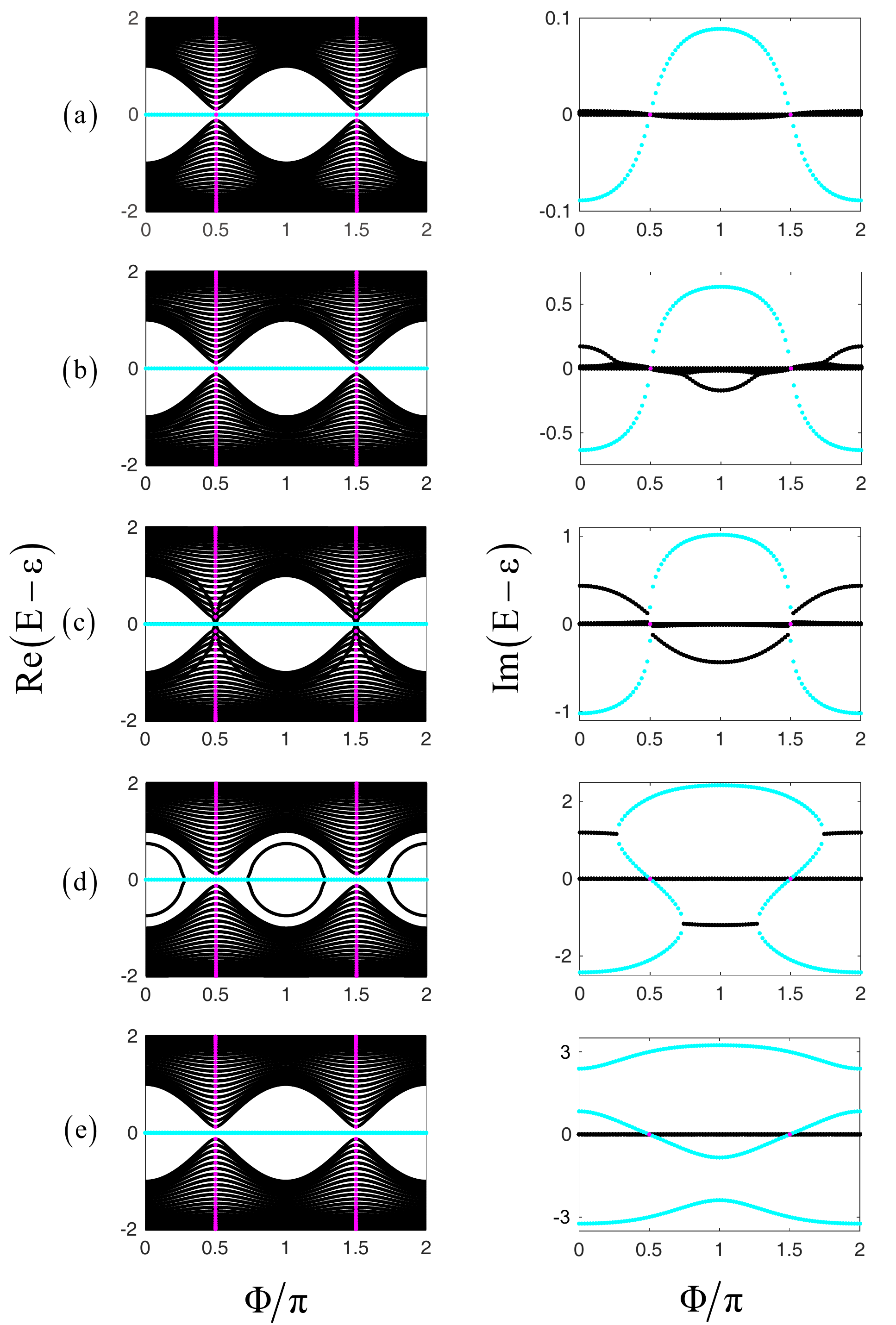}\label{fig6}
\centering
\caption{(Color online) The real and the imaginary parts of the energy eigenvalue spectrum of the $\mathcal{PT}$ asymmetric coupled cavity arrays model as functions of $\Phi$ with parameters $\delta=0.5$ and $\emph{N}=51$ for different $\kappa$. (a) $\kappa=0.1$, (b) $\kappa=0.7$, (c) $\kappa=1.1$, (d) $\kappa=2.5$, and (e) $\kappa=3.3$. Left and right figures represent the real and imaginary parts of the energy eigenvalue spectrum, respectively. The cyan points represent the real and imaginary parts of the purely imaginary energy eigenvalues and the magenta points represent the real and imaginary parts of the real energy eigenvalues.}
\end{figure}
\begin{figure}
\includegraphics[scale=0.6]{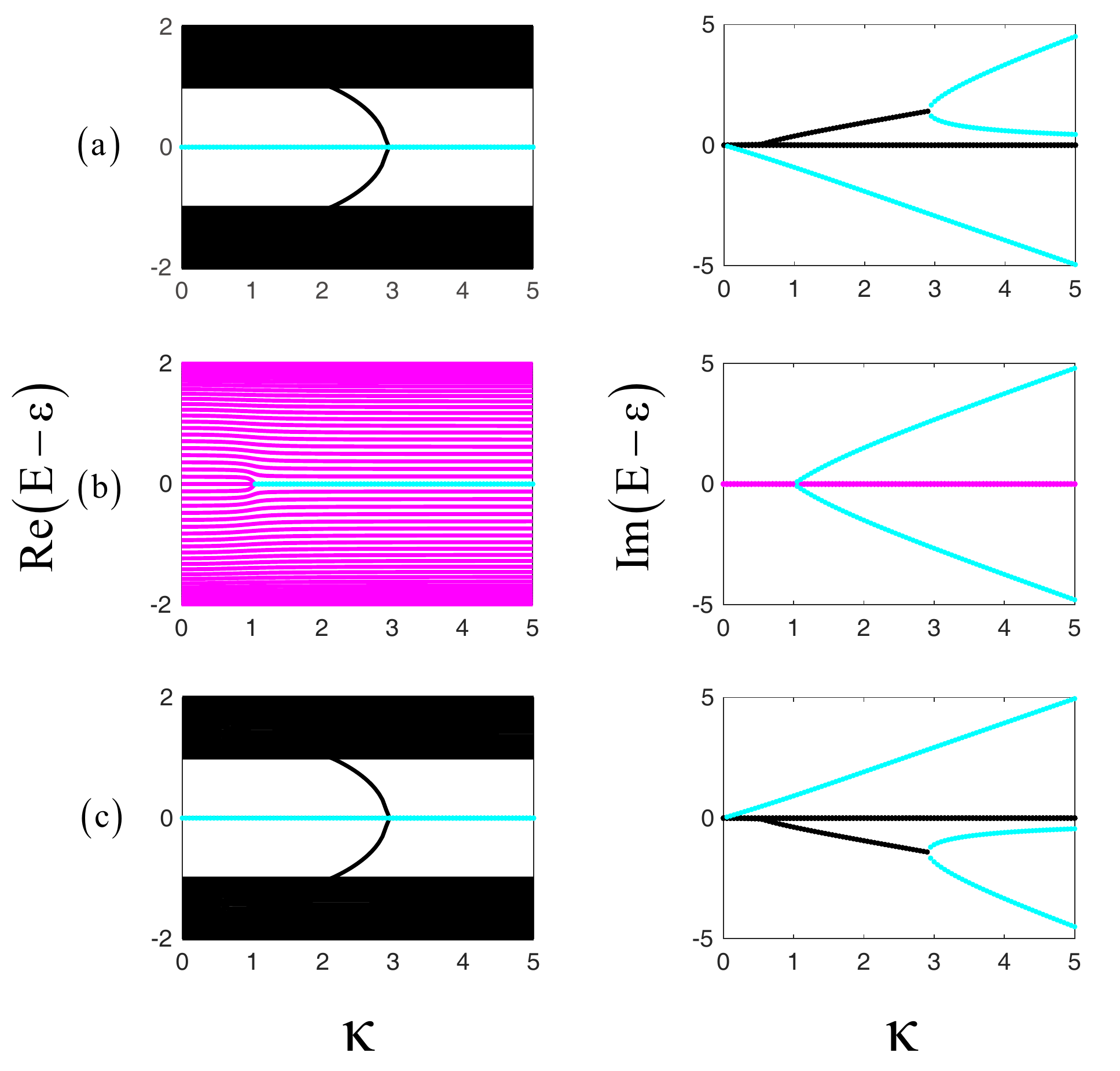}\label{fig7}
\centering
\caption{(Color online) The real and imaginary parts of the energy eigenvalue spectrum versus $\kappa$ for the system with different values of $\Phi$. (a) $\Phi=0$, (b) $\Phi=\frac{\pi}{2}$, and (c) $\Phi=\pi$. Other parameters are the same as Fig.~6. The cyan points represent the real and imaginary parts of the purely imaginary energy eigenvalues and the magenta points represent the real and imaginary parts of the real energy eigenvalues.}
\end{figure}

For studying the effects of the odevity of the number of cavities on the energy eigenvalue spectrum of the system, we also present the numerical results of the energy eigenvalue spectrum in the situation of odd number of cavities though the Hamiltonian of the system in this case is not $\mathcal{PT}$ symmetric, as shown in Fig.~6 and Fig.~7. For the system with a weak effective loss rate $\kappa=0.1$, as shown in Fig.~6(a), the complex energy eigenvalues composed of one purely imaginary energy eigenvalue and $N-1$ complex energy eigenvalues with weak imaginary parts emerge in the whole region of $\Phi$ except $\Phi=\frac{\pi}{2},\frac{3\pi}{2}$. On the contrary, there exists $\emph{N}$ real energy eigenvalues consisting of a single zero energy mode and $N-1$ real energy eigenvalues at $\Phi=\frac{\pi}{2},\frac{3\pi}{2}$. With $\kappa$ increasing, one of the absolute values of the imaginary parts belonging to the complex energy eigenvalues gradually increases, as shown in Fig.~6(b). In Fig.~6(c), one can see that a pair of purely imaginary energy eigenvalues with the form of $\pm ib$ emerge at $\Phi=\frac{\pi}{2},\frac{3\pi}{2}$ with $\kappa_{c^{\prime\prime}}(\frac{\pi}{2})=\kappa_{c^{\prime\prime}}(\frac{3\pi}{2})=1.01$. When further increasing $\kappa$, one of the imaginary parts belonging to the complex energy eigenvalues follows on a split when $\kappa>1.38$. At the same time, the other absolute values of the imaginary parts of the complex energy eigenvalues taper off, as shown in Figs.~6(d) and (e). Finally, the entire energy eigenvalue spectrum is made up of three purely imaginary energy eigenvalues and $N-3$ complex energy eigenvalues with weak imaginary parts except $\Phi=\frac{\pi}{2},\frac{3\pi}{2}$ which contains a single zero energy mode, a pair of purely imaginary energy eigenvalues, and $N-3$ real energy eigenvalues, as shown in Fig.~6(e). Specially, the energy eigenvalue spectrums versus $\kappa$ for the system with $\Phi=0,\frac{\pi}{2},\pi$ are given in Fig.~7, which is consist with the preceding numerical results. In the limit of $\kappa\rightarrow\infty$, one of the values of the three purely imaginary parts corresponding to the middle cyan line in Fig.~6(e) tends to zero, as shown in Figs.~7(a) and 7(c).

From the above figures, we can conclude that the difference revealed by the two systems due to odevity is noteworthy and the effects of the non-Hermitian terms on the $\mathcal{PT}$ symmetric and asymmetric systems are comparatively different in this case.

\subsection{The passive and active cavities at the second and penultimate positions}
\begin{figure}
\includegraphics[scale=0.7]{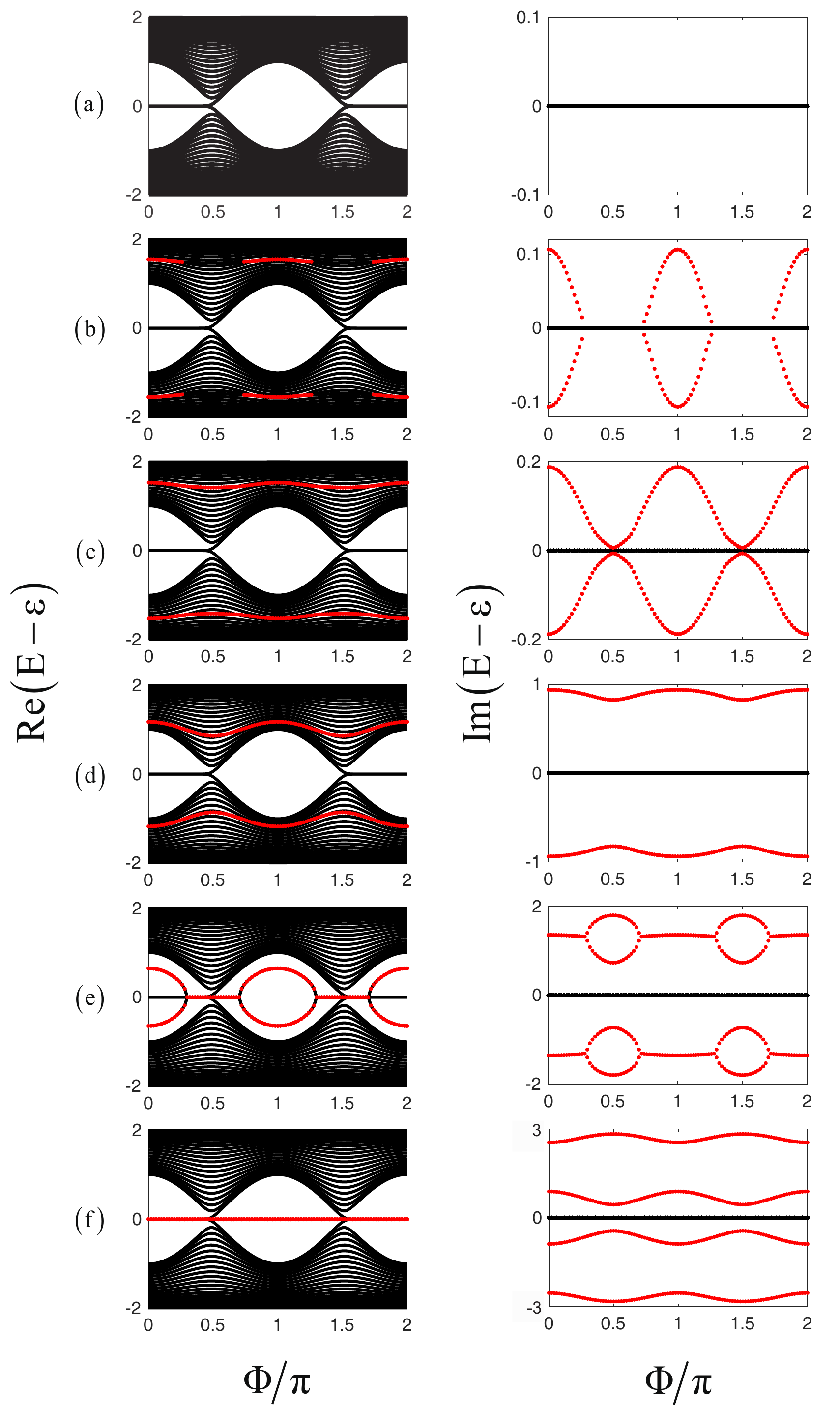}\label{fig8}
\centering
\caption{(Color online) The real and the imaginary parts of the energy eigenvalue spectrum of the $\mathcal{PT}$ symmetric coupled cavity arrays model as functions of $\Phi$ with parameters $\delta=0.5$ and $\emph{N}=50$ for different $\kappa$. (a) $\kappa=0.4$, (b) $\kappa=0.6$, (c) $\kappa=0.71$, (d) $\kappa=2$, (e) $\kappa=2.8$, and (f) $\kappa=3.5$. Left and right figures represent the real and imaginary parts of the energy eigenvalue spectrum, respectively. The red points represent the real and imaginary parts of the complex energy eigenvalues.}
\end{figure}
\begin{figure}
\includegraphics[scale=0.6]{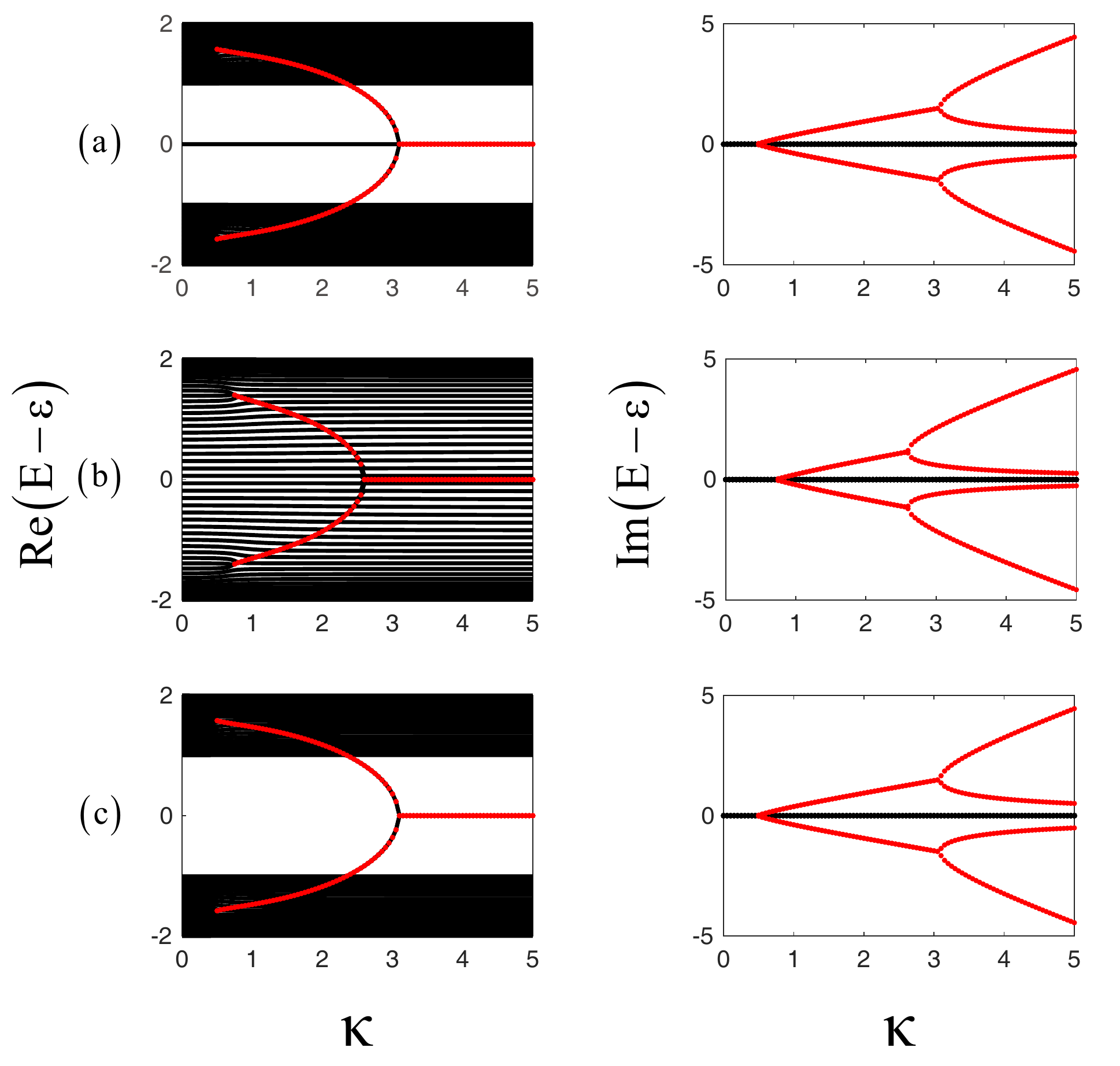}\label{fig9}
\centering
\caption{(Color online) The real and imaginary parts of the energy eigenvalue spectrum versus $\kappa$ for the system in different topological regimes. (a) System in the topologically nontrivial regime with $\Phi=0$, (b) system at the phase boundary points with $\Phi=\frac{\pi}{2}$, and (c) system in the topologically trivial regime with $\Phi=\pi$. Other
parameters are the same as Fig.~8. The red points represent the real and imaginary parts of the complex energy eigenvalues.}
\end{figure}
\begin{figure}
\centering
\includegraphics[scale=0.8]{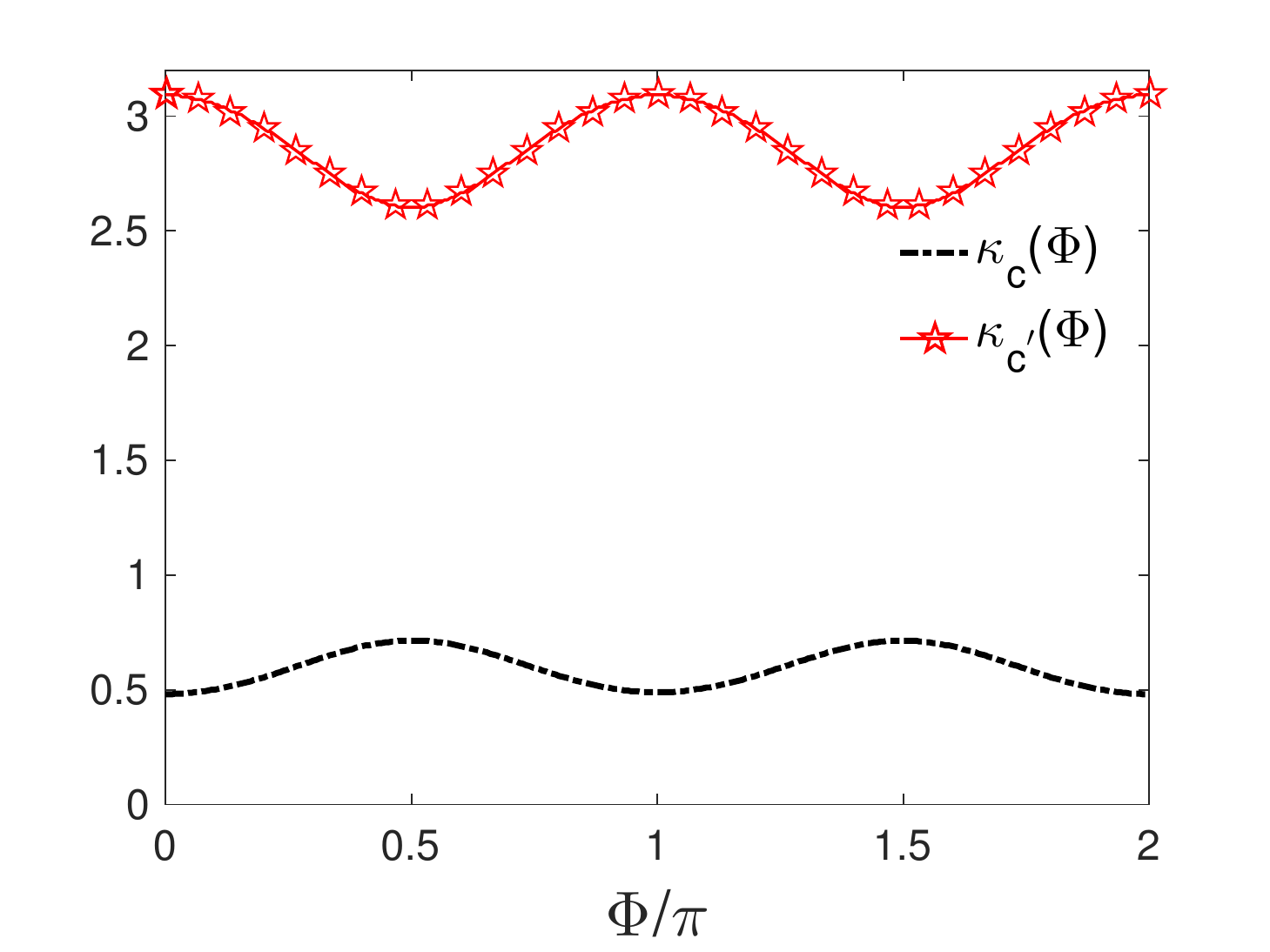}\label{fig10}
\caption{(Color online) The sketches of $\kappa_{c}(\Phi)$ and $\kappa_{c^{'}}(\Phi)$ for the system as function of $\Phi$.}
\end{figure}
\begin{figure}
\includegraphics[scale=0.7]{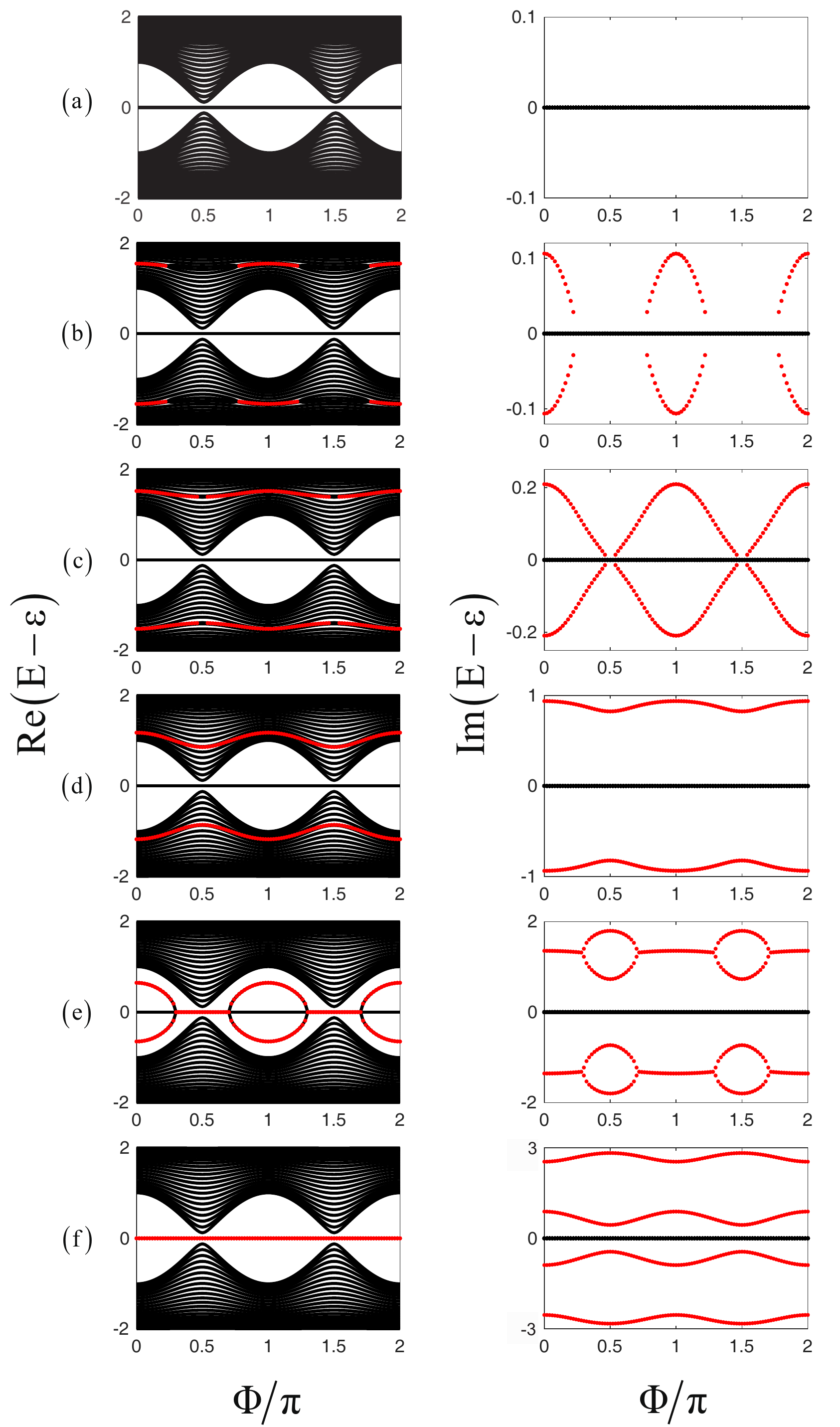}\label{fig11}
\centering
\caption{(Color online) The real and the imaginary parts of the energy eigenvalue spectrum of the $\mathcal{PT}$ asymmetric coupled cavities array model as a function of $\Phi$ with parameters $\delta=0.5$ and $\emph{N}=51$ for different $\kappa$. (a) $\kappa=0.4$, (b) $\kappa=0.6$, (c) $\kappa=0.74$, (d) $\kappa=2$, (e) $\kappa=2.8$, and (f) $\kappa=3.5$. Left and right figures represent the real and imaginary parts of the energy eigenvalue spectrum, respectively. The red points represent the real and imaginary parts of the complex energy eigenvalues.}
\end{figure}
\begin{figure}
\includegraphics[scale=0.6]{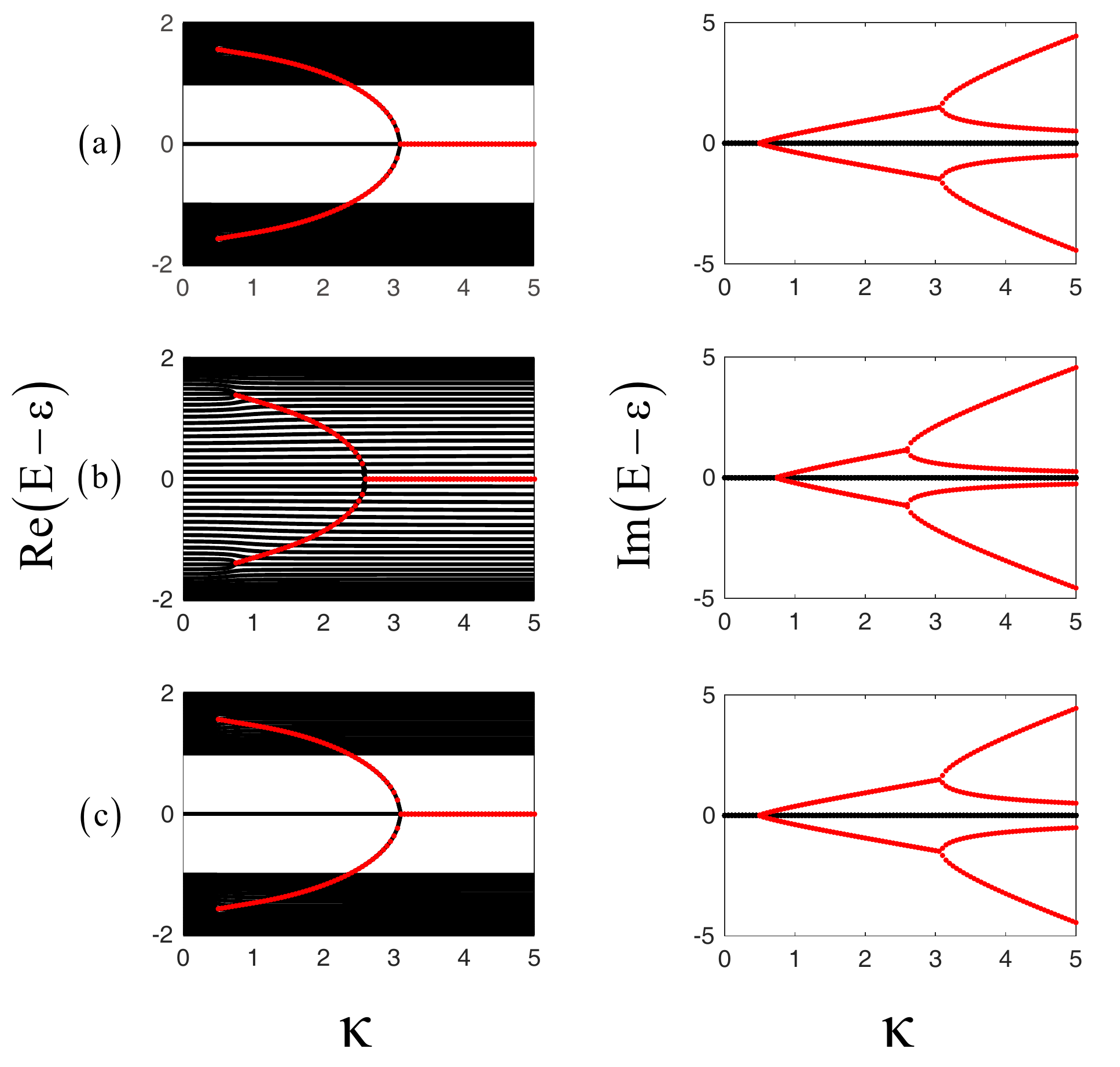}\label{fig12}
\centering
\caption{(Color online) The real and imaginary parts of the energy eigenvalue spectrum versus $\kappa$ for the system with different values of $\Phi$. (a) $\Phi=0$, (b) $\Phi=\frac{\pi}{2}$, and (c) $\Phi=\pi$. Other parameters are the same as Fig.~11. The red points represent the real and imaginary parts of the complex energy eigenvalues.}
\end{figure}

In this section, we consider that the passive and active cavities are placed at the second and penultimate positions in coupled cavities array. In this case, the Hamiltonian of the system becomes
\begin{eqnarray}\label{e07}
H_{\mathrm{s}}^\prime&=&\left[\sum_{j=2n-1}^{N-1}\left(1-\delta\cos\Phi\right)a_{j}^{\dag}a_{j+1}+\sum_{j=2n}^{N-1}\left(1+\delta\cos\Phi\right)a_{j}^{\dag}a_{j+1}
+\mathrm{H.c.}\right]\cr\cr
&&+\sum_{i=1}^{\emph{N}}\varepsilon a_{i}^{\dag}a_{i}-i\kappa a_{2}^{\dag}a_{2}+i\kappa a_{\emph{N-1}}^{\dag}a_{\emph{N-1}}.
\end{eqnarray}
Comparing with case $\mathrm{A}$, in the situation of even number of cavities, the $\mathcal{PT}$ symmetric system displays distinct behaviors in both the topologically nontrivial regime and phase boundary points, as shown in Fig.~8 and Fig.~9. One can clearly observe from Fig.~8(a) that there exists a real energy eigenvalue spectrum even $\kappa=0.4$. With $\kappa$ increasing, four complex energy eigenvalues with the form of $\pm a\pm ib$ emerge in the topologically nontrivial regime on the condition that $\kappa>\kappa_{c}(0)=0.474$. In the meantime, the spontaneous $\mathcal{PT}$ symmetry breaking transition is almost simultaneous in the both topologically nontrivial and trivial regimes, as shown in Fig.~8(b). Nevertheless, the spontaneous $\mathcal{PT}$ symmetry breaking lastly takes place at the phase boundary points as ever, as shown in Fig.~8(c). Then with further increasing $\kappa$, one can see from Fig.~8(d) that the values of the imaginary parts of the four complex energy eigenvalues increase markedly. More interestingly, when $\kappa>2.59$, the entire energy eigenvalue spectrum will undergo a second transition propagated from the phase boundary points to both sides with another critical value $\kappa_{c^{'}}(\Phi)$, corresponding to that four complex energy eigenvalues turn into two pairs of conjugated purely imaginary energy eigenvalues, as shown in Fig.~8(e). All of the phase regimes exhibit two pairs of conjugated purely imaginary energy eigenvalues when $\kappa>\kappa_{c^{'}}(0)=\kappa_{c^{'}}(\pi)=3.08$, as shown in Fig.~8(f). Particularly, we also present the real and imaginary parts of energy eigenvalues as functions of $\kappa$ for the system in different topological regimes by setting $\Phi=0,\frac{\pi}{2},\pi$, respectively, as shown in Fig.~9. A pair of conjugated purely imaginary energy eigenvalues also tend to zero in the limit of $\kappa\rightarrow\infty$ corresponding to the two central red lines in Fig.~8(f). Additionally, the sketch of $\kappa_{c}(\Phi)$ and $\kappa_{c^{'}}(\Phi)$ is also given in Fig.~10. In brief, the system can maintain $\mathcal{PT}$ symmetry when $\kappa\leq0.474$ corresponding to a real energy eigenvalue spectrum. A second transition occurs in both the topologically nontrivial and trivial regimes, definitely, so do the phase boundary points.

On the other hand, the numerical results of the energy eigenvalue spectrum for odd number of cavities are also given in Fig.~11 and Fig.~12. In spite of some subtle differences, such as Fig.~8(c) and Fig.~11(c), one can clearly see that the system still exhibits similar behaviors compared with the results for even number of cavities, viz., the complex energy eigenvalue spectrum shows the similar distribution behavior throughout the regime $\Phi\in[0,2\pi]$ for each of the same $\kappa$. The above phenomena clarify that the effects of the non-Hermitian terms on the $\mathcal{PT}$ symmetric and asymmetric systems are qualitatively same in this case.

\subsection{A sequence of the passive and active cavities}
\begin{figure}
\includegraphics[scale=0.7]{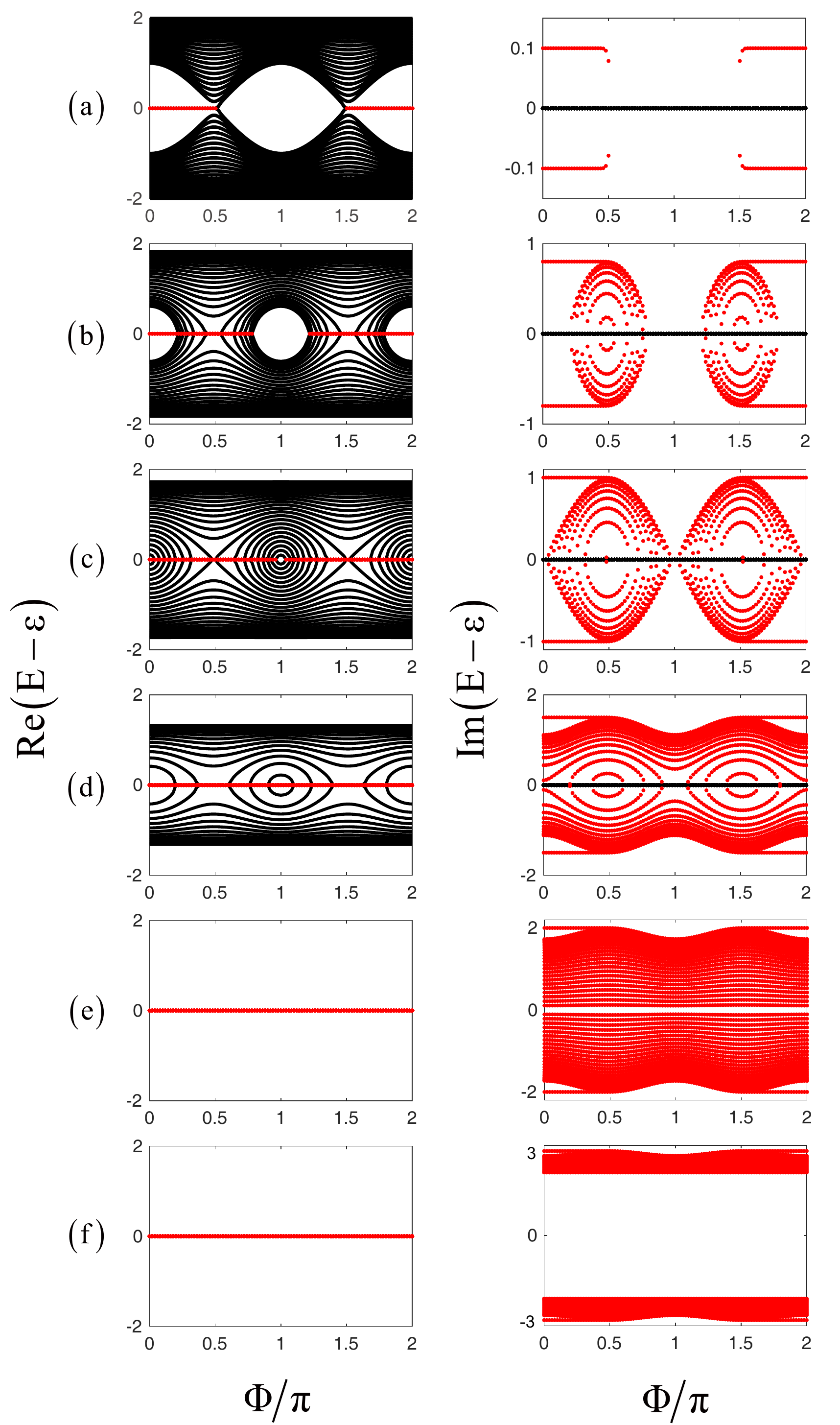}\label{fig13}
\centering
\caption{(Color online) The real and the imaginary parts of the energy eigenvalue spectrum of the $\mathcal{PT}$ symmetric coupled cavities array model as a function of $\Phi$ with parameters $\delta=0.5$ and $\emph{N}=50$ for different $\kappa$. (a) $\kappa=0.1$, (b) $\kappa=0.8$, (c) $\kappa=1$, (d) $\kappa=1.5$, (e) $\kappa=2$, and (f) $\kappa=3$. Left and right figures represent the real and imaginary parts of the energy eigenvalue spectrum, respectively. The red points represent the real and imaginary parts of the complex energy eigenvalues.}
\end{figure}

\begin{figure}
\includegraphics[scale=0.6]{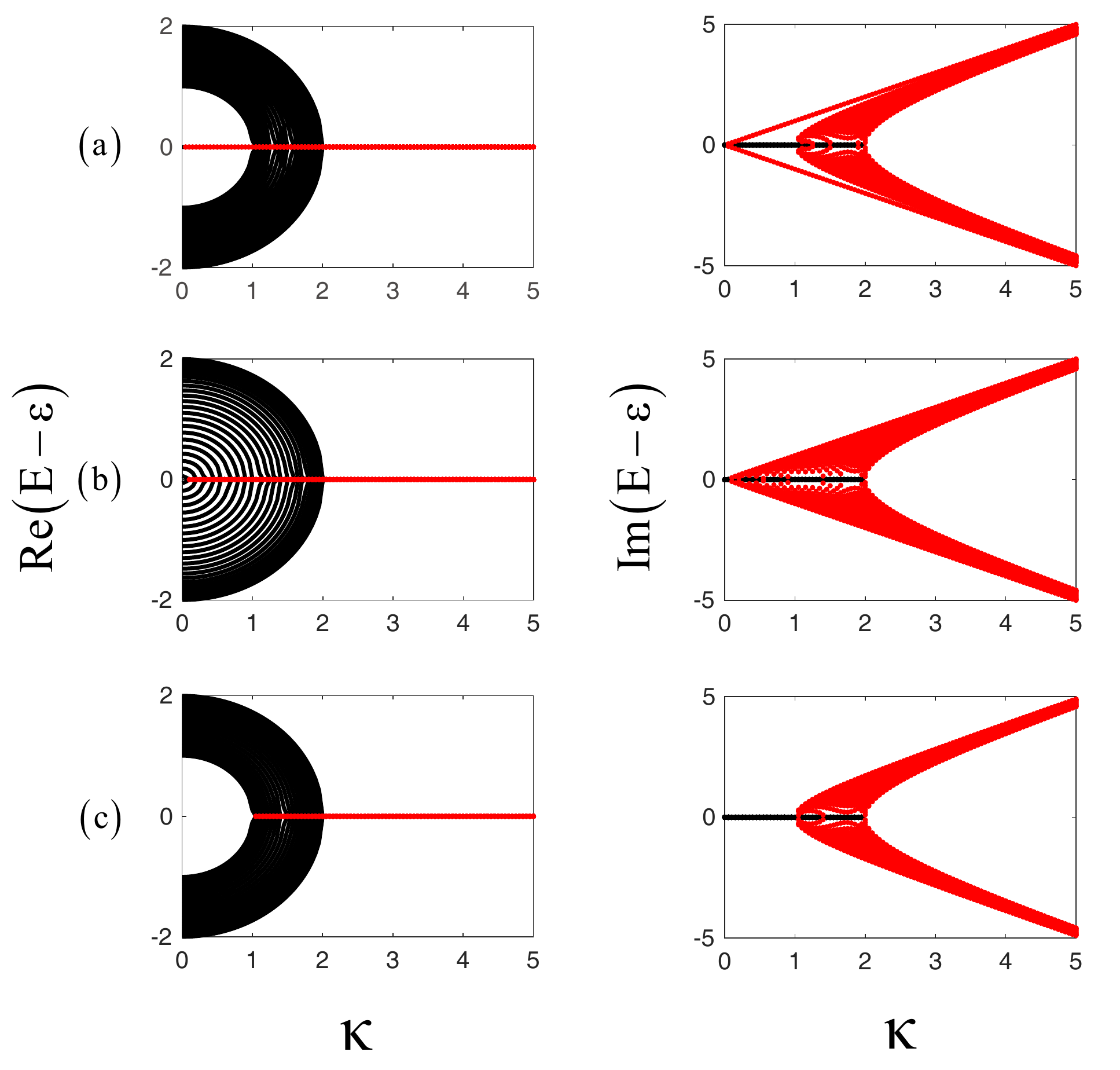}\label{fig14}
\centering
\caption{(Color online) The real and imaginary parts of the energy eigenvalue spectrum versus $\kappa$ for the system in different topological regimes. (a) System in the topologically nontrivial regime with $\Phi=0$, (b) system at the phase boundary points with $\Phi=\frac{\pi}{2}$, and (c) system in the topologically trivial regime with $\Phi=\pi$. Other
parameters are the same as Fig.~13. The red points represent the real and imaginary parts of the complex energy eigenvalues.}
\end{figure}

Now we turn to a more general $\mathcal{PT}$ symmetric case, i.e., the system is comprised of a sequence of the passive and active cavities. In this case the Hamiltonian of non-Hermitian coupled cavity arrays model becomes
\begin{eqnarray}\label{e07}
H_{\mathrm{s}}^{\prime\prime}&=&\sum_{j=2n-1}^{N-1}\left[1-\delta\cos\left(\Phi\right)\right]a_{j}^{\dag}a_{j+1}+\sum_{j=2n}^{N-1}\left[1+\delta\cos\left(\Phi\right)\right]a_{j}^{\dag}a_{j+1}
+\mathrm{H.c.}\cr\cr
&&+\sum_{i=2n-1}^{N}\left(\varepsilon-i\kappa\right)a_{i}^{\dag}a_{i}+
\sum_{i=2n}^{N}\left(\varepsilon+i\kappa\right)a_{i}^{\dag}a_{i},
\end{eqnarray}
where $\sum_{j=2n-1}^{N}$ ($\sum_{j=2n}^{N}$) denotes that the passive cavities (active cavities) are added at the odd (even) positions.

The numerical results of the energy eigenvalue spectrum are plotted in Fig.~13 and Fig.~14. In the case of weak effective loss rate $\kappa=0.1$, we find that a pair of conjugated purely imaginary energy eigenvalues with the form of $\pm ib$ emerge both in the topologically nontrivial regime and the phase boundary points, as shown in Fig.~13(a). With $\kappa$ increasing, the number of conjugated purely imaginary energy eigenvalues around the phase boundary points (including phase boundary points) begins to increase and the behavior will gradually diffuse to all phase regimes with further increasing $\kappa$, and the $\mathcal{PT}$ symmetry of the total system is spontaneously broken when $\kappa>1$, as shown in Figs.~13(b)$-$(d). Relatively, the number of the real energy eigenvalues decreases until there only exists a purely imaginary energy eigenvalue spectrum, as shown in Figs.~13(e)$-$(f). Figure~14 shows the real and imaginary parts of energy eigenvalue spectrum as functions of $\kappa$ for the system in different phase regimes by assigning $\Phi$ as $0,\frac{\pi}{2},\pi$, respectively. One can observe that in the topologically nontrivial regime it initially exhibits a pair of conjugated purely imaginary energy eigenvalues and sequently accompanies with the occurrence of large-number purely imaginary energy eigenvalues. With respect to the phase boundary points and the topologically trivial regime, there only exists the latter behavior. It is evident that the whole energy eigenvalue of the system becomes purely imaginary when $\kappa\geq2$.

Now we briefly analyze and discuss some practical issues in relation to the experimental realization of the proposed scheme. In order to connect to the promising candidates for an experimental realization, we refer our set-up to the photonic bandgap cavities in photonic crystals and the whispering-gallery microcavities. Experimentally, large-scale ultrahigh-Q coupled nanocavity arrays based on photonic crystals have been realized~\cite{M274108}. Additionally, the design and fabrication of active and passive two-dimensional photonic crystal devices based on GaAs/AlGaAs slab waveguide nano-resonators has been presented~\cite{T83182306}, which can be used as the set-up in our scheme. It was also shown in the laboratory that $\mathcal{PT}$ symmetry and $\mathcal{PT}$ symmetry breaking can be realized by utilizing the active and passive whispering-gallery microcavities. The experimental realization of $\mathcal{PT}$ symmetric optics on a chip in two directly coupled high-Q silica-microtoroid resonators with balanced effective gain and loss are mentioned in~\cite{31LXSCJLGGM852414} and it has also reported $\mathcal{PT}$ symmetry breaking in coupled optical resonators~\cite{32BSFFMGSFCL1039414}, which can be used for the active and passive cavities in our scheme. All of the above experimental constructions and progresses contribute to the experimental simulation of $\mathcal{PT}$ symmetric SSH model and are beneficial for experimental realization of the $\mathcal{PT}$ symmetric SSH model based on large-scale arrays of cavities.

\section{Conclusions}
In conclusion, we have proposed a scheme to study and simulate the $\mathcal{PT}$ symmetric SSH model based on non-Hermitian coupled cavities array model and investigated the effects of the position of the passive and active cavities on the spontaneous $\mathcal{PT}$ symmetry breaking behavior of the system. We divide it into three cases to analyze and discuss in detail, i.e., the passive and active cavities are located at, respectively, the two end positions, the second and penultimate positions, and each position in coupled cavities array. Furthermore, the odevity of the number of cavities is also considered to check the effects of the same non-Hermitian terms applied on the $\mathcal{PT}$ symmetric and asymmetric systems. In the first case, as to the situation of even number cavities, we find that the system exhibits different spontaneous $\mathcal{PT}$ symmetry breaking behaviors in the topologically nontrivial and trivial regimes. As long as the effective loss rate of the passive cavity $\kappa\neq0$, a pair of conjugated purely imaginary energy eigenvalues will emerge in the topologically nontrivial regime. While in the topological trivial regime, when $\kappa$ is smaller than the critical value $\kappa_c(\Phi)$, there is no spontaneous $\mathcal{PT}$ symmetry breaking behavior. However, the system exhibits the spontaneous $\mathcal{PT}$ symmetry breaking transition behavior with four complex energy eigenvalues and a second transition behavior with two pairs of conjugated purely imaginary energy eigenvalues at, respectively, critical values $\kappa_c(\Phi)$ and $\kappa_{c^{\prime}}(\Phi)$. For the odd number situation, the entire energy eigenvalue spectrum is composed of one purely imaginary energy eigenvalue and $N$-$1$ complex energy eigenvalues once $\kappa>0$ except $\Phi=\frac{\pi}{2}, \frac{3\pi}{2}$. For $\Phi=\frac{\pi}{2}, \frac{3\pi}{2}$, the system shows real energy eigenvalues which is composed of a single zero energy mode and $N-1$ real energy eigenvalues. However, if further increasing $\kappa$ and once up to critical value $\kappa_c(\frac{\pi}{2})=\kappa_c(\frac{3\pi}{3})=1.01$, the $N$ purely real energy eigenvalues begin to transform into a single zero energy mode, a pair of conjugated purely imaginary, and $N-3$ purely real energy eigenvalues. One of the imaginary parts of the complex energy eigenvalues also begin to split when $\kappa>1.38$ and at last the energy eigenvalue spectrum is composed of three purely imaginary energy eigenvalues and $N-3$ complex energy eigenvalues except for $\Phi=\frac{\pi}{2}, \frac{3\pi}{2}$. However, in the second case, as to the situation of even number cavities, we find that the system exhibits the same spontaneous $\mathcal{PT}$ symmetry breaking and second transition behaviors but different breaking degree in both the topologically nontrivial and trivial regimes and phase boundary points. For the situation of odd number of cavities, the system reveals similar behaviors compared with the situation of even number of cavities. In the third case, the system exhibits a pair of conjugated purely imaginary energy eigenvalues in both the topologically nontrivial regime (once $\kappa\neq0$) and phase boundary points for a weak $\kappa$. Additionally, with $\kappa$ further increasing, large-scale purely imaginary energy eigenvalues appear in the entire energy eigenvalue spectrum. At the end, all the energy eigenvalues become purely imaginary and there only exists a purely imaginary energy eigenvalue spectrum.

When the number of cavities is even, i.e., $\mathcal{PT}$ symmetric situation, one can clearly observe from the first and second cases that the effects of the position of the passive and active cavities on the spontaneous $\mathcal{PT}$ symmetry breaking behavior of the topologically nontrivial regime and phase boundary points are remarkable. Moreover, the results for the first case imply that the effects of the non-Hermitian terms on the $\mathcal{PT}$ symmetric and asymmetric systems due to the odevity are comparatively different. However, it means in the second case that the effects of the non-Hermitian terms on the $\mathcal{PT}$ symmetric and asymmetric systems due to the odevity are qualitatively same. We hope that the conclusions obtained in the present work will stimulate more interest in the study and simulation of the $\mathcal{PT}$ symmetric topological system or non-Hermitian topological system under the influences of environment based on arrays of cavities system.

\begin{center}
{\bf{ACKNOWLEDGMENTS}}
\end{center}

This work was supported by the National Natural Science Foundation of China under Grant Nos. 11465020, 11264042, 61465013, 11564041, and the Project of Jilin Science and Technology Development for Leading Talent of Science and Technology Innovation in Middle and Young and Team Project under Grant No. 20160519022JH.


\begin{thebibliography}{999}
\bibitem{001MC82304510}M. Z. Hasan and C. L. Kane, Rev. Mod. Phys. \textbf{82}, 3045 (2010).
\bibitem{002XS83105711}X. L. Qi and S. C. Zhang, Rev. Mod. Phys. \textbf{83}, 1057 (2011).
\bibitem{003JFGP83152311}J. Dalibard, F. Gerbier, G. Juzeli\={u}nas, and P. \"{O}hberg, Rev. Mod. Phys. \textbf{83}, 1523 (2011); N. Goldman, G. Juzeli\={u}nas, P. \"{O}hberg, and I. B. Spielman, Rep. Prog. Phys. \textbf{77}, 126401 (2014).
\bibitem{01YRKJI46262809}Y. J. Lin, R. L. Compton, K. Jim\'{e}nez-Garc\'{i}a, J. V. Porto, and I. B. Spielman, Nature (London) \textbf{462}, 628 (2009).
\bibitem{02MNEM303100513}M. S. Rudner, N. H. Lindner, E. Berg, and M. Levin, Phys. Rev. X \textbf{3}, 031005 (2013).
\bibitem{03KLRMAI10822530312}K. Jimen\'{e}z-Garc\'{i}a, L. J. LeBlanc, R. A.Williams, M. C. Beeler, A. R. Perry, and I. B. Spielman, Phys. Rev. Lett. \textbf{108}, 225303 (2012).
\bibitem{04FJDXRSL9006363814}F. Mei, J. B. You, D. W. Zhang, X. C. Yang, R. Fazio, S. L. Zhu, and L. C. Kwek, Phys. Rev. A \textbf{90}, 063638 (2014).
\bibitem{05AWYC1712500515}A. He, W. Luo, Y. Wang, and C. Gong, New J. Phys \textbf{17}, 125005 (2015).
\bibitem{06HMXX9205212215}H. Z. Shen, M. Qin, X. Q. Shao, and X. X. Yi, Phys. Rev. E \textbf{92}, 052122 (2015).
\bibitem{07ZHWX9303212016}Z. C. Shi, H. Z. Shen, W. Wang, and X. X. Yi, Phys. Rev. E \textbf{93}, 032120 (2016).
\bibitem{08HWX4645514}H. Z. Shen, W. Wang, and X. X. Yi, Sci. Rep. \textbf{4}, 6455 (2014).
\bibitem{W42169879}W. P. Su, J. R. Schrieffer, and A. J. Heeger, Solitons in Polyacetylene, Phys. Rev. Lett. \textbf{42}, 1698 (1979).
\bibitem{H21238880}H. Takayama, Y. R. Lin-Liu, and K. Maki, Phys. Rev. B \textbf{21}, 2388 (1980); W. P. Su, J. R. Schrieffer, and A. J. Heeger, ibid. \textbf{22}, 2099 (1980).
\bibitem{R133398}R. Jackiw and C. Rebbi, Phys. Rev. D \textbf{13}, 3398 (1976).
\bibitem{A6078188}A. J. Heeger, S. Kiverson, J. R. Schrieffer, and W. P. Su, Rev. Mod. Phys. \textbf{60}, 781 (1988).
\bibitem{J8818040102}J. Ruostekoski, G. V. Dunne, and J. Javanainen, Phys. Rev. Lett. \textbf{88}, 180401 (2002).
\bibitem{09PDG8419545211}P. Delplace, D. Ullmo, and G. Montambaux, Phys. Rev. B \textbf{84}, 195452 (2011).
\bibitem{10XEW4152313}X. Li, E. Zhao, and W. V. Liu, Nat. Commun. \textbf{4}, 1523 (2013).
\bibitem{11SKS11018040313}S. Ganeshan, K. Sun, and S. Das Sarma, Phys. Rev. Lett. \textbf{110}, 180403 (2013).
\bibitem{12I8615314}I. M. Georgescu, S. Ashhab, and Franco Nori, Rev. Mod. Phys. \textbf{86}, 153 (2014).
\bibitem{13M284906}M. J. Hartmann, F. G.S.L. Brand\~{a}o, and M. B. Plenio, Nat. Phys. \textbf{2}, 849 (2006).
\bibitem{14A285606}A. D. Greentree, C. Tahan, J. H. Cole, and L. C. L. Hollenberg, Nat. Phys. \textbf{2}, 856 (2006).
\bibitem{15F070200307}F. G.S.L. Brand\~{a}o, M. J. Hartmann, and M. B. Plenio, arXiv: quant-ph/0702003 (2007).
\bibitem{16D7603180507}D. G. Angelakis, M. F. Santos, and S. Bose, Phys. Rev. A \textbf{76}, 031805(R) (2007).
\bibitem{17M252708}M. J. Hartmann, F. G.S.L. Brand\~{a}o, and M. B. Plenio, Laser Photonics Reviews \textbf{2}, 527 (2008).
\bibitem{01R94}R. Shankar, Princples of Quantum Mechanics (Springer, New York, 1994).
\bibitem{02CS80524398}C. M. Bender and S. Boettcher, Phys. Rev. Lett. \textbf{80}, 5243 (1998).
\bibitem{03CDH7002500104}C. M. Bender, D. C. Brody, and H. F. Jones, Phys. Rev. D \textbf{70}, 025001 (2004); H. F. Jones, J. Phys. A \textbf{39}, 10123 (2006).
\bibitem{04I4215300109}I. Rotter, J. Phys. A \textbf{42}, 153001 (2009).
\bibitem{05IB80289798}I. Y. Goldsheid and B. A. Khoruzhenko, Phys. Rev. Lett. \textbf{80}, 2897 (1998).
\bibitem{06J6316510801}J. Heinrichs, Phys. Rev. B \textbf{63}, 165108 (2001).
\bibitem{07L4226520409}L. G. Molinari, J. Phys. A \textbf{42}, 265204 (2009).
\bibitem{08SUN10108040208}S. Klaiman, U. Gunther, and N. Moiseyev, Phys. Rev. Lett. \textbf{101}, 080402 (2008).
\bibitem{09AZY8204381810}A. A. Sukhorukov, Z. Xu, and Y. S. Kivshar, Phys. Rev. A \textbf{82}, 043818 (2010).
\bibitem{10HDVIT10903390212}H. Ramezani, D. N. Christodoulides, V. Kovanis, I. Vitebskiy, and T. Kottos, Phys. Rev. Lett. \textbf{109}, 033902 (2012).
\bibitem{11S10312360109} S. Longhi, Phys. Rev. Lett. \textbf{103}, 123601 (2009).
\bibitem{12ZKRD10003040208}Z. H. Musslimani, K. G. Makris, R. El-Ganainy, and D. N. Christodoulides, Phys. Rev. Lett. \textbf{100}, 030402 (2008).
\bibitem{13XJHXQYC11024390213}X. B. Luo, J. H. Huang, H. H. Zhong, X. Z. Qin, Q. T. Xie, Y. S. Kivshar, and C. H. Lee, Phys. Rev. Lett. \textbf{110}, 243902 (2013).
\bibitem{14YT8415310111}Y. C. Hu and T. L. Hughes, Phys. Rev. B \textbf{84}, 153101 (2011).
\bibitem{15BRS8906210214}B. Zhu, R. Lu, and S. Chen, Phys. Rev. A \textbf{89}, 062102 (2014).
\bibitem{16XTYP9201211615}X. Wang, T. Liu, Y. Xiong, and P. Tong, Phys. Rev. A \textbf{92}, 012116 (2015).
\bibitem{17QBSLR9402211916}Q. B. Zeng, B. Zhu, S. Chen, L. You, and R. Lu, Phys. Rev. A \textbf{94}, 022119 (2016).
\bibitem{18ORTB10303040209}O. Bendix, R. Fleischmann, T. Kottos, and B. Shapiro, Phys. Rev. Lett. \textbf{103}, 030402 (2009).
\bibitem{19LZ8005210709}L. Jin and Z. Song, Phys. Rev. A \textbf{80}, 052107 (2009).
\bibitem{20YDMA8203010310}Y. N. Joglekar, D. Scott, M. Babbey, and A. Saxena, Phys. Rev. A \textbf{82}, 030103(R) (2010); Y. N. Joglekar and A. Saxena, ibid. \textbf{83}, 050101(R) (2011); D. D. Scott and Y. N. Joglekar, ibid. \textbf{83}, 050102(R) (2011); S. Longhi, ibid. \textbf{88}, 052102 (2013).
\bibitem{21LZ8103210910}L. Jin and Z. Song, Phys. Rev. A \textbf{81}, 032109 (2010); W. H. Hu, L. Jin, Y. Li, and Z. Song, ibid. \textbf{86}, 042110 (2012); L. Jin and Z. Song, Commun. Theor. Phys. \textbf{54}, 73 (2010).
\bibitem{22G8205240410}G. L. Giorgi, Phys. Rev. B \textbf{82}, 052404 (2010).
\bibitem{23CGXZ9001210314}C. Li, G. Zhang, X. Z. Zhang, and Z. Song, Phys. Rev. A \textbf{90}, 012103 (2014).
\bibitem{24RKDZ32263207}R. El-Ganainy, K. G. Makris, D. N. Christodoulides, and Z. H. Musslimani, Opt. Lett. \textbf{32}, 2632 (2007).
\bibitem{25KRDZ10010390408} K. G. Makris, R. El-Ganainy, D. N. Christodoulides, and Z. H. Musslimani, Phys. Rev. Lett. \textbf{100}, 103904 (2008).
\bibitem{26CKRDMD619210}C. E. Ruter, K. G. Makris, R. El-Ganainy, D. N. Christodoulides, M. Segev, and D. Kip, Nat. Phys. \textbf{6}, 192 (2010).
\bibitem{27AGDRMVGD10309390209}A. Guo, G. J. Salamo, D. Duchesne, R. Morandotti, M. Volatier-Ravat, V. Aimez, G. A. Siviloglou, and D. N. Christodoulides, Phys. Rev. Lett. \textbf{103}, 093902 (2009).
\bibitem{28SBUHMAF10802410112}S. Bittner, B. Dietz, U. Gunther, H. L. Harney, M. Miski-Oglu, A. Richter, and F. Schafer, Phys. Rev. Lett. \textbf{108}, 024101 (2012)
\bibitem{29ACMGDU48816712}A. Regensburger, C. Bersch, M. Ali Miri, G. Onishchukov, D. N. Christodoulides, and U. Peschel, Nature (London) \textbf{488}, 167 (2012).
\bibitem{N2011}N. Moiseyev, Non-Hermitian Quantum Mechanics (Cambridge University Press, Cambridge, England, 2011).
\bibitem{30JRY9205383715}J. Li, R. Yu, and Y. Wu, Phys. Rev. A \textbf{92}, 053837 (2015).
\bibitem{31LXSCJLGGM852414}L. Chang, X. Jiang, S. Hua, C. Yang, J. Wen, L. Jiang, G. Li, G. Wang, and M. Xiao, Nat. Photon. \textbf{8}, 524 (2014).
\bibitem{32BSFFMGSFCL1039414}B. Peng, \d{S}. K. \"{O}zdemir, F. Lei, F. Monifi, M. Gianfreda, G. Long, S. Fan, F. Nori, C. M. Bender, and L. Yang, Nat. Phys. \textbf{10}, 394 (2014).
\bibitem{M274108}M. Notomi, E. Kuramochi, and T. Tanabe, Nat. Photon. \textbf{2}, 741 (2008).
\bibitem{T83182306}T. Stomeo, V. Errico, A. Salhi, A. Passaseo, R. Cingolani, A.D'Orazio, M. De Sario, V. Marrocco, V. Petruzzelli, F.Prudenzano, and M. De Vittorio, Microelectron. Eng. \textbf{83}, 1823 (2006).
\end{thebibliography}
\end{document}